\documentclass[aps,twocolumn,superscriptaddress,preprintnumbers,showpacs,article,prl,longbibliography]{revtex4-1}

\usepackage[latin9]{inputenc}
\usepackage{color}
\usepackage{verbatim}
\usepackage{amsmath}
\usepackage{amssymb}
\usepackage{graphicx}
\usepackage{esint}
\usepackage{esvect}
\usepackage{braket}
\usepackage{amsmath,bm}
\usepackage{graphicx}
\usepackage{subfigure}
\usepackage{url}
\usepackage{lipsum}
\usepackage{lipsum}
\usepackage[dvipsnames]{xcolor}
\usepackage{hyperref}
\usepackage[resetlabels]{multibib}

\makeatletter
\@ifundefined{textcolor}{}
{%
 \definecolor{BLACK}{gray}{0}
 \definecolor{WHITE}{gray}{1}
 \definecolor{RED}{rgb}{1,0,0}
 \definecolor{GREEN}{rgb}{0,1,0}
 \definecolor{BLUE}{rgb}{0,0,1}
 \definecolor{CYAN}{cmyk}{1,0,0,0}
 \definecolor{MAGENTA}{cmyk}{0,1,0,0}
 \definecolor{YELLOW}{cmyk}{0,0,1,0}
}

\usepackage{mathrsfs}

\draft
\makeatother
\begin{document}

\title{Minimal one-dimensional model of bad metal behavior from fast particle-hole scattering}
\author{Yan-Qi Wang}
\affiliation{Department of Physics, University of California,  Berkeley, California 94720, USA}
\affiliation{Materials Sciences Division, Lawrence Berkeley National Laboratory, Berkeley, California 94720, USA}
\author{Roman Rausch}
\affiliation{Technische Universit\"at Braunschweig, Institut f\"ur Mathematische Physik, Mendelssohnstra\ss e 3, 38106 Braunschweig, Germany}
\author{Christoph Karrasch}
\affiliation{Technische Universit\"at Braunschweig, Institut f\"ur Mathematische Physik, Mendelssohnstra\ss e 3, 38106 Braunschweig, Germany}
\author{Joel E. Moore}
\affiliation{Department of Physics, University of California,  Berkeley, California 94720, USA}
\affiliation{Materials Sciences Division, Lawrence Berkeley National Laboratory, Berkeley, California 94720, USA}

\begin{abstract}
A strongly interacting plasma of linearly dispersing electron and hole excitations in two spatial dimensions (2D), also known as a Dirac fluid, can be captured by relativistic hydrodynamics and shares many universal features with other quantum critical systems.  We propose a one-dimensional (1D) model to capture key aspects of the 2D Dirac fluid while including lattice effects and being amenable to non-perturbative computation.  When interactions are added to the Dirac-like 1D dispersion without opening a gap, we show that this kind of irrelevant interaction is able to preserve Fermi-liquid-like quasi-particle features while relaxing a zero-momentum charge current via collisions between particle-hole excitations, leading to resistivity that is linear in temperature via a mechanism previously discussed for large-diameter metallic carbon nanotubes. We further provide a microscopic lattice model and obtain numerical results via density-matrix renormalization group (DMRG) simulations, which support the above physical picture. The limits on such fast relaxation at strong coupling are of considerable interest because of the ubiquity of bad metals in experiments.
\end{abstract}

\maketitle

{\it Introduction.--} A strongly interacting plasma of linearly dispersing electron and hole excitations in two dimension, also known as a Dirac fluid, shares many universal features with other quantum critical systems. With particle-hole symmetry preserved, under external electric field, there exists a ``zero momentum mode'' in the Dirac fluid which carries a non-vanishing charge current~\cite{Hartnoll2007,Muller20081,Muller20082}: electrons and holes move symmetrically in opposite directions. Protected by conservation of momentum, in a continuous translationally invariant system, such a charge current could only be relaxed via scattering within the quasi-particles in the current.  The most studied example of this kind of Dirac fluid is the electron-hole plasma in high mobility graphene at the charge neutrality point, which is believed to have Planckian-bounded dissipation~\cite{Hartnoll2007,Muller20081,Muller20082,Kimgraphene,Lucas2016A,Lucas2016B,Phan2013,Sun2016,Sun3285,Lucas_2018,Gallagher158,Fritz2008,Ku2020}, refering to a relaxation or scattering time $\tau_p \sim \hbar/k_B T$ set only by temperature and the Planck constant~\cite{Zaanen2004,Zaane2019}.  There is considerable experimental evidence for the importance of such relaxation rates as an upper bound in a broad range of ``bad metals''~\cite{Orenstein1990,Bruin2013,Varma_2020,Else2021}, most famously in the linear-in-temperature resistivity of some cuprate superconductors at optimal doping, in contrast to the standard form $\rho = \rho_0 + A T^2$ of Fermi liquids.   While the origin of linear-in-temperature resistivity in the normal state of high $T_c$ superconductors at optimal doping remains an open question~\cite{Takagi1992,Zaanen2004,Zaane2019,Varma_2020,Orenstein1990,Bruin2013,Else2021}, hydrodynamic studies for quantum critical fluids suggest one kind of answer~\cite{Hartnoll2007,Ku2020,Muller20081,Muller20082,Fritz2008,Zaanen2004,Gallagher158,Nam2017,Stewart2001,Zaane2019}: a quantum critical electron fluid with maximal Planckian dissipation is one theoretical route to linear-$T$ resistivity, even if the nature of a quantum critical point near optimal doping is difficult to probe because of the intervening superconductivity.

Conceptually, if one were to take a sheet of graphene and wrap it into a metallic armchair nanotube, one might expect some signs of 2D Dirac fluid transport along the tube axis to be preserved.  Indeed, Balents and Fisher argued that interactions in a sufficiently large nanotube, while expected ultimately to open a gap, might show a linear-in-$T$ resistivity over a range of temperatures, based on particle-hole scattering as a perturbation~\cite{Balents_1997}. As nothing in the Dirac fluid picture is manifestly specific to two dimensions, one could ask whether similar features could be obtained in one spatial dimension, where metallic transport is well known to have unique features~\cite{Bertini_2021}. On the other hand, previous models have been studied to explore whether it is possible to relax the current in an impurity-free, non-integrable 1D system at finite temperature, but these generally have parametrically slower relaxation than required for linear-in-$T$ resistivity~\cite{Samokhin_1998,Matveev2013,Huang2013,Bulchandani12713}.  All these encourage us to look elsewhere for a 1D model which can support Planckian dissipation and linear-$T$ resistivity, in analogy with the Dirac fluid. As more non-perturbative calculations are available in one dimension both theoretically and numerically, constructing a 1D Dirac fluid and increasing interactions to strong coupling is a test of one origin of Planckian dissipation.

In this letter, we propose a 1D model with no observable gap, and use a kinetic theory approach to determine its resistivity~\cite{Damle1997,Sachdev1997,Sachdev1998,Landau2013,Kadanoff1963,Maxime2020}. To check that this physics can be realized in a solid, we then introduce a microscopic lattice model that manifests the aforementioned internal scattering process. We further use time-dependent density-matrix renormalization-group (DMRG) simulations~\cite{White_1992,Vidal2003,Vidal2007,Schollowock2011} to confirm the gaplessness of the lattice model, and compute the current relaxation at finite temperature.

\begin{figure}[!h]
\centering 
\includegraphics[width=1\columnwidth]{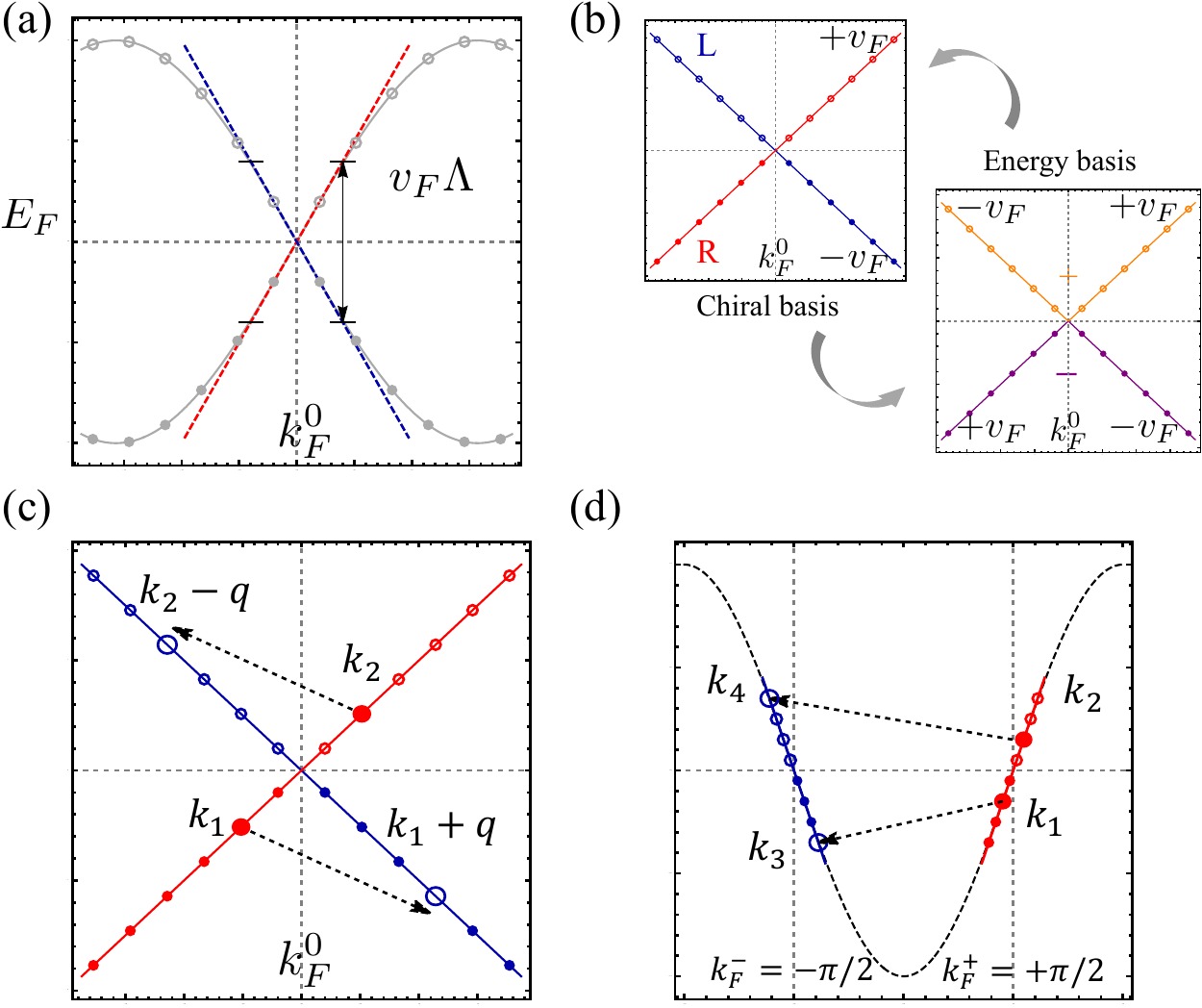}
\caption{\label{Illustration} Fermionic spectrum linearization and scattering processes based on it. (a) A 1D non-interacting metallic band can be linearized close to the Fermi level $E_F$ within the energy cutoff $v_F \Lambda$.  Note that the left fermi point and right fermi point coincides at $k_F^0$. (b) The linearized spectrum can be described by the chiral basis, with the red modes moving to the right with velocity $+v_F$ and blue modes moving to the left with velocity $-v_F$ in real space. The linearized spectrum can also be written in the energy basis, where states are labeled by positive (orange) and negative (purple) energies. The velocities for different quasi-particles are also shown in the figure.  (c) Illustration for particle-hole scattering or fast Umklapp-like scattering (FUS). Two right movers with momentum $k_1$ and $k_2$ are scattered to the left moving branch with momentum $k_1 + q$ and $k_2 -q$. Note that the total momentum is conserved in this process. (d) Conventional Umklapp scattering in 1D metal for two right movers ($k_1,k_2$) scattering into two left movers ($k_3,k_4$). Note that for conventional Umklapp scattering, the Fermi points for the right and left movers are different (say at $k_F^\pm = \pm \pi/2$.). The momentum is conserved only up to a reciprocal vector $G = 2\pi$, i.e., there is a large momentum transfer in the scattering process.}
\end{figure}

{\it Continuous model. --} The low-energy theory of a non-interacting 1D metal can be obtained by linearizing the spectrum near the Fermi level.  When the Fermi points for the left- and right-moving linear branches coincide with each other, we arrive at a Dirac-like crossing, as shown in Fig.~[\ref{Illustration}.(a)]. The linearized free Hamiltonian around the Fermi point, $H_0$, can be written in a chiral basis as
\begin{equation}\label{ChiralBasis}
	H_0 = v_F \int \frac{dk}{2\pi} k[\psi^\dagger_R(k) \psi_R(k) -  \psi^\dagger_L(k)\psi_L(k)]
\end{equation}
where $\psi_R(k)$ and $\psi_L(k)$ stand for the annihilation operators for right- and left-moving chiral fermion modes at one-dimensional momentum $k$, respectively, and $v_F$ is the Fermi velocity near the Fermi level. The above chiral basis can be transformed into the energy basis~\cite{Fritz2008,Sachdev1998}, in which $\gamma_+(k)$ and $\gamma_-(k)$ annihilate an electron with energy above and below the Dirac node, respectively:
\begin{equation}
	\begin{pmatrix}
		\gamma_+(k) \\
		\gamma_-(k)
	\end{pmatrix}
	=\frac{1}{2}\begin{pmatrix}
		1 + \vartheta(k) & 1 - \vartheta(k) \\
		1 - \vartheta(k) & 1 + \vartheta(k)
	\end{pmatrix}
	\begin{pmatrix}
		\psi_R(k) \\
		\psi_L(k)
	\end{pmatrix},
\end{equation}
where $\vartheta(x) =1$ for $x >0$ and $\vartheta(x) = -1$ for $x<0$. Note that the density of states vanishes at the Dirac node, so hereafter we can neglect the singularity at $k=0$ itself. With this, the free Hamiltonian is transformed into the following form:
\begin{equation}\label{FreeModelEnergyBasis}
	H_0 = v_F \int \frac{dk}{2\pi}|k| [\gamma^\dagger_+(k)\gamma_+(k) -\gamma^\dagger_-(k) \gamma_-(k)].
\end{equation}
Both chiral and energy basis are plotted in Fig.[\ref{Illustration}.(b)].

Now we study the full Hamiltonian $H$ with an interaction $H_{\rm int}$ turned on:
\begin{equation}\label{Model}
	H = H_0 + H_{\rm int}.
\end{equation}
We would like the interaction to introduce the following fast Umklapp-like scattering (FUS) among the chiral fermions~\footnote{Since there are different terminologies appearing in the literature, note that particle-hole scattering can also be viewed as a kind of two-particle Umklapp scattering but with no loss of momentum, as explained below Figure 1.}:
\begin{widetext}
\begin{equation}\label{InteractionChiralBasis}
\begin{aligned}
	H_{\rm int} &= \int \frac{dk_1}{2\pi} \frac{dk_2}{2\pi} \frac{dq}{2\pi} V(q)[\psi^\dagger_R(k_1 + q) \psi^\dagger_R(k_2-q) \psi_L(k_2) \psi_L(k_1) + \psi^\dagger_L(k_1 + q) \psi^\dagger_L(k_2-q)\psi_R(k_2) \psi_R(k_1)].
\end{aligned}
\end{equation}
\end{widetext}
This process takes two electrons on the same branch to the opposite branch, as shown in Fig.~[\ref{Illustration}.(c)]~\cite{Xu2006,Wu2006,Fradkinbook,ShankarBook}. Unlike the conventional Umklapp scattering for a 1-component model (see in Fig.~[\ref{Illustration}.(d)]), the FUS defined here does not carry large momentum transfer, as the left- and right-moving branches' Fermi points coincide at the Dirac node.  One can alternately view one of the processes as the scattering of a hole rather than an electron.  We will see in a particle-hole symmetric system, a current of oppositely directed particles and holes can have zero total momentum, allowing the current to relax through momentum-conserving collisions.

The interaction Eq.~[\ref{InteractionChiralBasis}] can also be written in the energy basis:
\begin{equation}\label{InteractionEnergyBasis}
\begin{aligned}
	H_{\rm int} &=\sum_{\lambda_1\lambda_2\lambda_3 \lambda_4} \int \frac{dk_1}{2\pi} \frac{dk_2}{2\pi} \frac{dq}{2\pi}  T_{\lambda_1\lambda_2\lambda_3\lambda_4} (k_1,k_2,q) \\
	&\times \gamma^\dagger_{\lambda_4}(k_1 + q) \gamma^\dagger_{\lambda_3}(k_2-q) \gamma_{\lambda_2}(k_2) \gamma_{\lambda_1}(k_1),
\end{aligned}
\end{equation}
where the $\lambda_{1,...,4}$ in the summation take the value of $\pm$ and the structure factor $T_{\lambda_1\lambda_2\lambda_3\lambda_4} (k_1,k_2,q) = T^2_{\lambda_1\lambda_2\lambda_3\lambda_4} + T^3_{\lambda_1\lambda_2\lambda_3\lambda_4}$ where: $		T^2_{\lambda_1\lambda_2\lambda_3\lambda_4} = V(q)[\lambda_1 \vartheta(k_1) - \lambda_4 \vartheta(k_1 + q)] [\lambda_2 \vartheta(k_2) - \lambda_3 \vartheta(k_2-q)]/16$ and $	T^3_{\lambda_1\lambda_2\lambda_3\lambda_4} =V(q)[ 1 - \lambda_1\lambda_3  \vartheta(k_1^2+qk_1)][ 1 - \lambda_2\lambda_3 \vartheta(k_2^2-qk_2)]/16$ are the matrices which indicate the scattering amplitudes among electrons with positive and negative energy.

{\it Kinetic theory.--} One can use the kinetic (hydrodynamic) theory to describe transport properties~\cite{Damle1997,Sachdev1997,Sachdev1998,Landau2013,Kadanoff1963}. Note that, for the particle density $\rho(x) = \psi^\dagger_R(x) \psi_R(x) + \psi^\dagger_L(x) \psi_L(x)$, the continuity equation $\partial_t \rho(x) + \partial_x j(x) = 0$ gives the $U(1)$ current density $j(x) =v_F[ \psi^\dagger_R(x)\psi_R(x) - \psi^\dagger_L(x)\psi_L(x)]$. Assume the charge carried by each particle (hole) is $+Q$ ($-Q$), the total charge current $J$ reads~\cite{Fritz2008,Sachdev1998}:
\begin{subequations}\label{TotalCurrent}
\begin{align}
	J&=  v_F Q\sum_{r = R,L} \int \frac{dk}{2\pi} r\psi^\dagger_r(k) \psi_r(k) \label{CurrentChiralBasis} \\
	&= v_F Q\sum_{\lambda = \pm} \int \frac{dk}{2\pi} \frac{\lambda k}{|k|} \gamma^\dagger_\lambda(k)\gamma_\lambda(k). \label{CurrentEnergyBasis}
\end{align}
\end{subequations} 
Note that in Eq.[\ref{CurrentChiralBasis}], we have  $r = 1$ for right movers ($R$) and $r =-1$ for left movers ($L$).
Similarly, the total momentum reads:
\begin{equation}\label{TotalMomentum}
	P = \sum_{r = R,L} \int \frac{dk}{2\pi} k \psi^\dagger_r(k) \psi_r(k) = \sum_\lambda \int \frac{dk}{2\pi} k \gamma^\dagger_\lambda(k) \gamma_\lambda(k).
\end{equation}

 We define the distribution functions under the energy basis $\gamma_\pm$ at time $t$ as:
\begin{equation}
	f_\lambda(k,t) = \langle \gamma^\dagger_\lambda(k,t) \gamma_\lambda(k,t) \rangle.
\end{equation}
In the equilibrium, without external perturbation, these are related to the Fermi distribution function $f^0(p) $, such that $f_\pm(k,t) = f^0(\pm \epsilon_k)=[{e^{ (\pm \epsilon_k -\mu)/k_BT}+ 1}]^{-1}$ with $\mu$ the chemical potential. From standard bosonization~\cite{Supp}, one can safely assume that turning on the interaction will neither open a gap, nor have an immediate modification of single-particle spectrum of Eq.~[\ref{FreeModelEnergyBasis}]. Then the Fermi liquid picture survives, $\epsilon_\lambda(k) =\lambda v_F|k|$, with $\lambda = \pm$ for two flavors of quasiparticles: the positive energy ones with the distribution function $f_{+}(k,t)$, and the negative energy ones with the distribution function $f_-(k,t)$.

\begin{figure}[!h]
\centering 
\includegraphics[width=1\columnwidth]{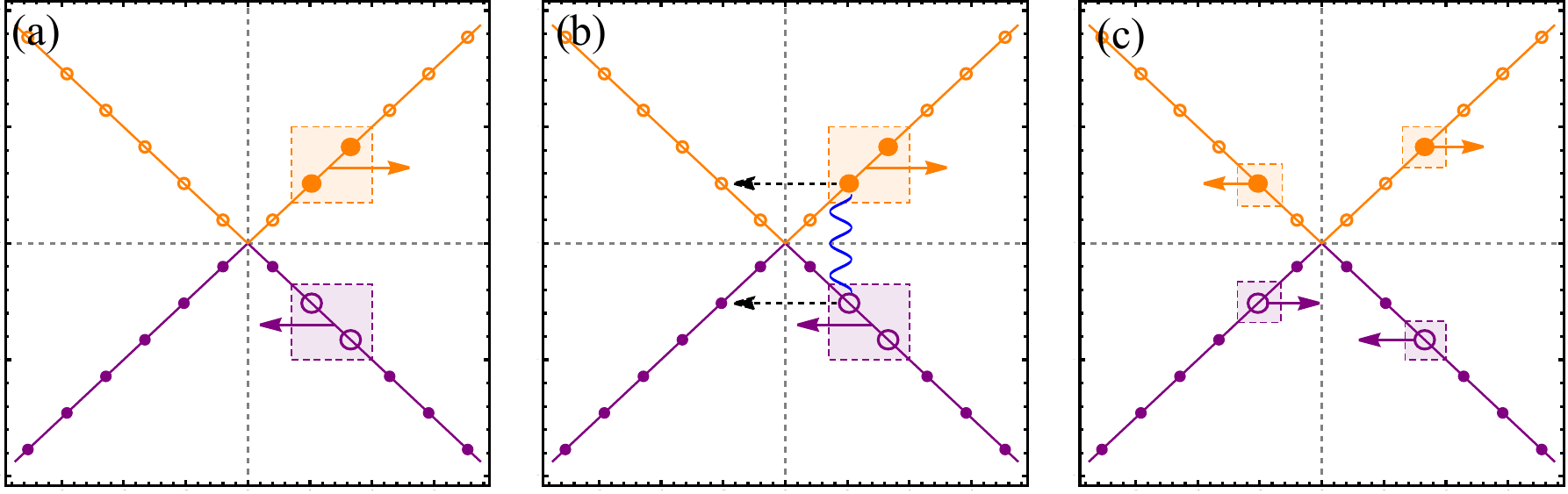}
\caption{\label{CurrentPlot} Generation and relaxation of charge current in a particle-hole symmetric system. (a) Generation of the zero momentum mode under external electric field. The net charge current for the states in the plot is $J =4Qv_F$. (b) Collision between the particle and hole via the interaction (waved line) based on the initial state shown in (a). (c) Final state after scattering process in (b), which has zero momentum and zero charge current. }
\end{figure}

The quantum Boltzmann equation with collisions reads:~\cite{Fritz2008,Arnold_2000,Sachdev1998}
\begin{equation}\label{QBECollision}
	\begin{aligned}
		\bigg{[} \frac{\partial}{\partial t} + Q {E}(t)  \frac{\partial}{\partial { k}} \bigg{]}f_\lambda({ k}, t)  &= -\frac{2\pi}{v_F}  \int \frac{dk_1}{2\pi } \frac{dq }{2\pi} {\mathcal R}.
	\end{aligned}
\end{equation}
The integrand ${\mathcal R} = {\mathcal R}_1 + {\mathcal R}_2$~\footnote{Note that we only have two integrals and one delta function on the right hand side of Eq.~[\ref{QBECollision}], as the conservation of energy and conservation of momentum have the same requirement for linear dispersion in one dimension. }, capturing the scattering among excitations, can be derived by a simple application of Fermi's golden rule together with the interaction in Eq.~[\ref{InteractionEnergyBasis}]. The first part is the scattering among different flavors of excitations (particle-hole to particle-hole) ${\mathcal R}_1 = \delta[{(|k|-|{k_1}|)}-{(|{{k} + { q}}| - |{{k}_1 -{q}})}]R_1({k},{k}_1,{q}) \{ f_\lambda({ k},t) f_{-\lambda}({ k}_1,t)[1-f_\lambda({ k}+{ q},t)][1-f_{-\lambda}({k}_1-{q},t)]- [1-f_\lambda({ k},t)][1- f_{-\lambda}({ k}_1,t)]f_\lambda({ k}+{ q},t)f_{-\lambda}({ k}_1-{ q},t)]\} $, with scattering amplitude ${R}_{1}({ k},{ k}_1,{ q}) = 4|T_{+--+}({ k},{k}_1,{q}) - T_{+-+-}({k},{k}_1,-{k}-{q} + { k}_1)|^2$. The second part ${\mathcal R}_2$ captures the scattering among same flavor of excitations (particles to particles or holes to holes), which will not contribute to the resistivity at leading order~\footnote{This is in accordance with the fact that the 1D chiral fluid can not be relaxed in the absence of impurities, see~\cite{Supp}}. To solve Eq.~[\ref{QBECollision}], we first parametrize the change in $f_\lambda$ from its equilibrium value by using the ansatz~\cite{Fritz2008,Arnold_2000}:
\begin{equation}\label{Ansatz}
\begin{aligned}
	f_\lambda({k},\omega) &= 2\pi \delta(\omega)f^0(\lambda \epsilon_k) +Q \frac{kE(\omega)}{|k|} f^{0}(\lambda \epsilon_k)\\
	&\times [1-f^0(\lambda \epsilon_k)]g_\lambda(\epsilon_k,\omega),
\end{aligned}
\end{equation}
with $g_\lambda(\epsilon_k,\omega)$ a function to be determined, and we have replaced $f_\lambda(k,t)$ with its Fourier counterpart in frequency domain $f_\lambda(k,\omega)$. When $\mu =0$, the system is at particle-hole symmetric point, and an applied electric field $E(\omega)$ generates deviations in the distribution functions for particles and holes with opposite signs. This is due to the fact that the driving term Eq.~[\ref{QBECollision}] is odd under $\lambda \rightarrow -\lambda$, thus the deviation also has to be asymmetric in $\lambda$: $g_\lambda(\epsilon_k,\omega) = \lambda g(k,\omega)$. In coordinate space, there will be newly generated holes (particles) moving align (anti-align) with the external electric field. This can be viewed as the generation of particle-hole pairs. For the states within the orange and purple square shown in the Fig.~[\ref{CurrentPlot}.(a)], at the same $k$ point, the particle and hole has opposite momentum, and each particle-hole pair has zero total momentum defined by Eq.~[\ref{TotalMomentum}] in the presence of particle-hole symmetry. On the other hand, since the particles and holes carry opposite charge, if they move in the opposite directions, the total current given by Eq.~[\ref{TotalCurrent}] is non-zero. Substituting Eq.~[\ref{Ansatz}] into Eq.~[\ref{QBECollision}], one could derive a solution for $g$ via the variational methods~\cite{Fritz2008,Arnold_2000,Supp}. Combined with Eq.~[\ref{CurrentEnergyBasis}], with the definition of charge conductivity $\sigma = J/E$, we arrive at: 
\begin{equation}\label{CollisionConductivity}
	\begin{aligned}
		\sigma(\omega) &= \frac{\langle J \rangle}{E(\omega)}  \approx \frac{2Q^2}{h} \frac{\hbar v_F}{-i\hbar \omega + \kappa k_BT},
	\end{aligned}
\end{equation} 	
where $\kappa$ is associated with inter flavor scattering:
\begin{equation}\label{NumericalCoefficient}
\kappa =\int \frac{d\tilde k}{2\pi} \frac{d \tilde q}{2\pi} \frac{4R_1(\tilde k, -\tilde k,\tilde q)/v_F^2}{(e^{-|\tilde k|} +1) (e^{|\tilde k|} +1)(e^{|\tilde k+\tilde q|} +1)(e^{-|\tilde k + \tilde q|} +1)}.
\end{equation}

As a check for the validity of our kinetic theory, we first notice that in the collionless limit ${\mathcal R}={\mathcal R}_1 = {\mathcal R}_2= 0$ such that $\kappa = 0$, we shall see:
\begin{equation}\label{FreeConductivityMain}
	\sigma(\omega) \approx \frac{2Q^2}{h} \frac{\hbar v_F}{-i\hbar \omega + \eta},
\end{equation}
with $\eta$ a a positive infinitesimal. This is consistent with the bosonization results for clean system in $1$D~\cite{Giamarchi2003quantum}. The presence of a Drude peak in the low-frequency limit is the signature of ballistic transport~\cite{Giamarchi2003quantum,Sirker2009}.

In the presence of FUS, $\kappa \neq 0$. Compared with the low frequency diverging result for the collisionless case in Eq.~[\ref{FreeConductivityMain}], the conductivity with collisions has some broadening at finite temperature. This shows that the zero momentum mode can be relaxed solely by the momentum conserved internal scattering process among excitations. Such physical picture is plotted in Fig.~[\ref{CurrentPlot}.(b-c)]. From Eq.~[\ref{CollisionConductivity}], we find the resistivity $\rho = 1/\sigma$ in the DC limit has a linear-$T$ dependence, i.e., the Planckian dissipation:
\begin{equation}
	\rho(\omega \rightarrow 0)\sim AT,
\end{equation}
with the coefficient $A = \pi \kappa k_B/ Q^2 v_F$. A one-dimensional Dirac system whose linear dispersion survives the interaction can be captured by a single model-dependent parameter, the Fermi velocity $v_F$. Combined with the temperature $T$, the only time scale in the continuous limit is the Planckian time $\tau_p = \hbar/k_BT$. Such a time scale gives the scattering rate for particle-hole excitations in an impurity-free Dirac system, and sets up an upper bound for the resistivity at finite temperature $\rho =A T$. The coefficient $A \propto |V(q)/v_F|^2$, which shows that the resistivity is also positively related to the interaction strength in the perturbative region, in accordance with the previous results in wrapping graphene sheet to large-diameter metallic carbon nanotubes~\cite{Balents_1997}.

{\it Lattice model.--}We start from a fermionic lattice model which possesses a low-energy Hamiltonian Eq.~[\ref{Model}], reads:
\begin{equation}\label{NewModel}
	\begin{aligned}
	   \tilde H &= \tilde H_0 + \tilde  H_2 + \tilde H_3 \\
		\tilde H_0 &= +t \sum_i [-i\xi a_i^\dagger b_i + i \xi b_i^\dagger a_i -ib_i^\dagger a_{i+1} + ia^\dagger_{i+1}b_i] \\
		 \tilde H_2 &= \frac{V_2}{4}{ \sum_i\big{[}  (a_i^\dagger b_i - b^\dagger_i a_i)(a^\dagger_{i+1}b_{i+1}-b^\dagger_{i+1}a_{i+1})} \\
		&+ { (b^\dagger_i a_{i+1} - a^\dagger_{i+1}b_i)(b^\dagger_{i+1}a_{i+2} -a^\dagger_{i+2}b_{i+1})  }\big{]} \\
		\tilde H_3 &= \frac{V_3}{4} { \sum_i\big{[} (a^\dagger_i a_i -b^\dagger_i b_i)(a^\dagger_{i+1}a_{i+1}-b^\dagger_{i+1}b_{i+1}) } \\
		&+ {  (b_i^\dagger b_i -a^\dagger_{i+1}a_{i+1})(b^\dagger_{i+1}b_{i+1}-a^\dagger_{i+2}a_{i+2}) \big{]}}.
	\end{aligned}
\end{equation}
Here, the $\tilde H_0$ stands for the free Hamiltonian and $\tilde H_2 + \tilde H_3$ is the interaction. The kinetic part $\tilde H_0$ can be connected to the integrable XX model~\cite{Bahovadinov2019}. The addition of $\tilde H_2 + \tilde H_3$ breaks the integrability (see a plot of the crossover of level statistics from Poisson to Wigner-Dyson in the supplementary materials~\cite{Supp,Bruus1997}), which justify the legitimacy of using kinetic equations in our analytic calculations. The form of the interaction is obtained by seeking to construct a Hamiltonian which has FUS as its naive continuum limit, then symmetrizing the Hamiltonian, i.e., ensuring that it does not include a relevant, gap-opening dimerization at least at leading order.  The $a^\dagger_i$ and $b^\dagger_i$ denote the creation operators for two distinct degrees of freedom at the same point in $i$-th unit-cell. When $\xi =1$, the Bloch Hamiltonian for $\tilde H_0$ can be linearized around $k =0$, and has the Dirac-like structure as given in Eq.~[\ref{ChiralBasis}]. When $V_2= V_3$, to the leading order the interaction will only contain the FUS given in Eq.~[\ref{InteractionChiralBasis}]. We further provide the charge current density operator $j_{i+1} = (+tQ)[b_i^\dagger a_{i+1} + a^\dagger_{i+1} b_i] $
which with density-density interactions satisfies the standard continuity equation $ {\partial \rho_i(t)}/{\partial t} + (j_{i+1}- j_i) =0$ for the local charge density of each unit cell $\rho_n = Q(a^\dagger_n a_n + b^\dagger_n b_n)$~\cite{Supp}. The DC conductivity may be found via Kubo formula:
\begin{equation}
    \sigma = \lim_{t_M \rightarrow \infty} \lim_{N\rightarrow \infty} \frac{1}{NT} {\rm Re} \int_0^{t_M} dt\langle J(t) J(0) \rangle,
    \label{eq:sigma1D}
\end{equation}
where the total charge current for a system with $N$ unit cells at time $t$ is $J(t) = \sum_{i=1}^N j_i(t)$~\cite{Bulchandani12713,Kubo1957a,Kubo1957b,Luttinger1964,Kapustin2019}.

{\it DMRG results.--} We first confirm that the lattice model Eq.~[\ref{NewModel}] is gapless for $V_2,V_3< 4t$ using the density matrix renormalization group~\cite{White_1992,Supp}. To evaluate the conductivity, we proceed using standard techniques~\cite{Feiguin_2005,Karrasch_2013,Barthel_2016}: Finite temperatures are implemented by going from a pure state to the density operator. We enlarge the local Hilbert space to include an auxiliary part, which is traced out when performing expectation values. The state at $\beta = 1/T = 0$ is exactly initialized on a finite chain with $L=96$ and then propagated to the desired $\beta$. After that, the state is perturbed by applying the current operator and propagated in real time up, which yields the current correlation function $\langle J(t)J\rangle/L$. We have chosen the system size large enough as not to allow the current to reach the finite system boundaries at the end of the simulation. Since the Hamiltonian contains more than nearest-neighbor interactions, we cannot use a standard time-evolving block decimation algorithm but instead employ the time-dependent variational principle (TDVP)~\cite{Haegeman_2016}. We use a two-site TDVP algorithm and variationally compress the MPS at each time step~\cite{Supp,Hubig_2017,Kennes_2016}.

\begin{figure}[th]
\centering 
\includegraphics[width=1\columnwidth]{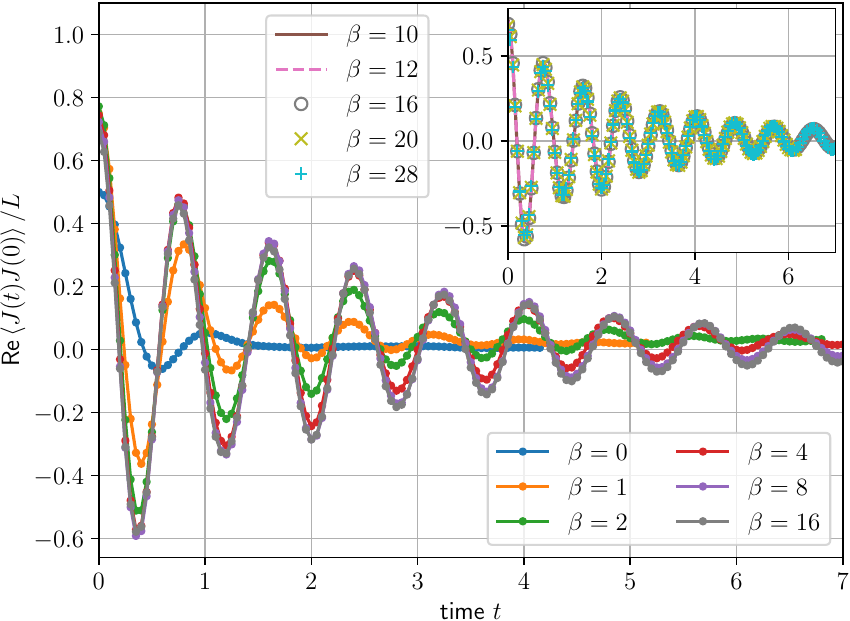}
\caption{\label{fig:DMRGcurrent}Time-dependent current-current correlation function at $V_2=V_3=3$ calculated using DMRG.}
\end{figure}

For small $V=V_2=V_3$, the entanglement buildup is relatively small, but we are also very close to the integrable point $V=0$, resulting in very long relaxation times; and vice versa for large $V$. Faced with these trade-offs, we find that we need to go to $V=3$ to be able to evaluate the current correlations. The results are shown in Fig.~\ref{fig:DMRGcurrent}. It turns out that we still cannot reach time scales which are long enough to quantitatively compute the integral in Eq.~(\ref{eq:sigma1D}), we observe that below $T=1/8\sim1/10$, the different curves essentially collapse onto one curve for the times we are able to access. Assuming that this collapse will continue to hold for the inaccessible times as well, this means that the integral over $\langle J(t)J\rangle/L$ becomes independent of $T$ in this regime. Due to the prefactor of $1/T$ in Eq.~(\ref{eq:sigma1D}), this points towards a resistivity which is indeed proportional to $T$ in the low-temperature regime. The seemingly complicated model Eq.~[\ref{NewModel}] provides a route to realizing the conjectured Planckian upper bound to the resistivity for a class of realistic interacting semimetals in 1D with local and non-random interactions~\cite{Balents_1997,Yoshioka1999}.

{\it Conclusion.--}  We proposed a model for 1D Dirac fermionic system as well as its lattice counterpart, and showed that quasi-particles broaden from collisions at finite temperature compared with the well known diverging results for 1D two-channel ballistic transport.  Verifying transport similar to that proposed for the 2D Dirac liquid in a 1D model provides an alternative point of view on the origin of bad metallic behavior, and observing the dominance of Umklapp-like scattering in our model complements other possibilities for transport theory in one dimension dominated by other irrelevant operators~\cite{Rice2017}. The analytical and numerical methods available to explore transport in low spatial dimensions make it feasible to search for evidence of other physics originally proposed for higher dimensions, as we have done here for the Dirac fluid~\cite{syk,Lieb1963,Hild2014,Tang2018,Edwin2019}.

{\it Acknowledgment.--} This work was supported as part of the Center for Novel Pathways to Quantum Coherence in Materials, an Energy Frontier Research Center funded by the U.S. Department of Energy, Office of Science, Basic Energy Sciences (Y.-Q.W. and J.E.M.). R.R. and C.K. acknowledge support by the Deutsche Forschungsgemeinschaft (DFG, German Research Foundation) through the Emmy Noether program (KA3360/2-1). Y.-Q. W thanks Vir Bulchandani and Jyong-Hao Chen for the early introductions to transport theory. J.E.M. acknowledges support from a Simons Investigatorship and thanks Chandra Varma for helpful discussions.

\bibliography{apssamp2023v1}

\clearpage{}

\noindent

\onecolumngrid

\pagebreak
\widetext
\begin{center}
\textbf{\large Supplemental Materials for ``Minimal one-dimensional model of bad metal behavior from fast particle-hole scattering''}
\end{center}

\renewcommand\theequation{S\arabic{equation}}
\renewcommand\thefigure{S\arabic{figure}}
\setcounter{equation}{0}
\setcounter{figure}{0}
\section{Motivation}
A strongly interacting plasma of linearly dispersing electron and hole excitations in two dimension, also known as a Dirac fluid, shares many universal features with other quantum critical systems. With particle-hole symmetry preserved, under external electric field, there exists a ``zero momentum mode'' in the Dirac fluid which carries a non-vanishing charge current~\cite{Hartnoll2007,Muller20081,Muller20082}: electrons and holes move symmetrically in opposite directions. Protected by conservation of momentum, in a continuous translationally invariant system, such a charge current could only be relaxed via scattering within the quasi-particles in the current.  The most studied example of this kind of Dirac fluid is the electron-hole plasma in high mobility graphene at the charge neutrality point, which is believed to have Planckian-bounded dissipation~\cite{Kimgraphene,Hartnoll2007,Muller20081,Muller20082,Lucas2016A,Lucas2016B,Phan2013,Sun2016,Sun3285,Lucas_2018,Gallagher158,Fritz2008,Ku2020}.

Here Planckian dissipation refers to a relaxation or scattering time $\tau_p \sim \hbar/k_B T$ set only by temperature and the Planck constant~\cite{Zaanen2004,Zaane2019}.  There is considerable experimental evidence for the importance of such relaxation rates as an upper bound in a broad range of ``bad metals''~\cite{Orenstein1990,Bruin2013,Varma_2020}, most famously in the linear-in-temperature resistivity of some cuprate superconductors at optimal doping, in contrast to the standard form $\rho = \rho_0 + A T^2$ of Fermi liquids.  (Note that there can be mechanisms of dissipation that involve the Planckian time scale but do not lead to changes in resistivity, for example in translation-invariant systems where current is conserved by the dissipation process.)  The general mechanism of Planckian dissipation remains contested.  As nothing in the Dirac fluid picture is manifestly specific to two dimensions, one could ask whether similar features could be obtained in one spatial dimension, where metallic transport is well known to have unique features~\cite{Bertini_2021}. Because more non-perturbative calculations are available in one dimension both theoretically and numerically, constructing a one-dimensional Dirac fluid and increasing interactions to strong coupling is a test of one origin of Planckian dissipation.

Conceptually, if one were to take a sheet of graphene and wrap it into a metallic armchair nanotube, one might expect some signs of 2D Dirac fluid transport along the tube axis to be preserved.  Indeed, Balents and Fisher argued that interactions in a sufficiently large nanotube, while expected ultimately to open a gap, might show a linear-in-$T$ resistivity over a range of temperatures, based on particle-hole scattering as a perturbation~\cite{Balents_1997}. (Since there are different terminologies appearing in the literature, note that particle-hole scattering can also be viewed as a kind of two-particle Umklapp scattering but with no loss of momentum, as explained below Fig.~[MainText.1]). Our goal here is to find a single-chain model with no observable gap, use a kinetic theory approach to determine its resistivity beyond perturbation theory, and then verify the bad metal regime by taking advantage of the remarkable progress in dynamical matrix product state calculations.

%~\footnote{Here and after, for simplicity, all the Fig.~[num] and Eq.~[num] without the prefactor ``S'' refers to corresponding figures and equations presented in the main text.}

Of course, metals in one dimension are generally more sensitive to electron-electron interactions than in higher dimensions, resulting in a Luttinger liquid rather than the Fermi liquid familiar from higher dimensions.  Since the Luttinger liquid also starts from a Fermi surface with isolated points, and its existence is by now well established, the existence of alternative Dirac-fluid physics in one dimension must depend on the details of a microscopic model.  The past few years have seen a renaissance in the dynamics of one-dimensional models, including, even without disorder and the possibility of localization, a new kind of hydrodynamics in integrable models resulting from the inhibition of relaxation from extra conservation laws~\cite{Bertini_2021}.  Planckian dissipation is expected to appear in the opposite limit, where relaxation is happening as rapidly as possible.

The goal of this work is to understand whether one-dimensional Dirac fluids can be engineered in realistic lattice models, how they relate to known physics such as the Luttinger liquid, and whether they can show the fast relaxation that underlies Planckian behavior.  Throughout we use the word ``fast'' to indicate that the scattering process induces the equilibrium found from kinetic theory before any higher-order interaction effect opens a gap; it is difficult to rule out either analytically or numerically the emergence of a tiny energy gap, which will affect dynamics only at the longest time scales and lowest temperatures.

In the remainder of this introduction, we review when effective hydrodynamic descriptions appear in metallic materials and some existing mechanisms or examples of Planckian dissipation.  Real solids have at most discrete translational symmetries, which can be broken down by impurities.  Under some conditions, the momentum relaxation processes like Umklapp and impurity scattering possess a characteristic time scale $\tau_{r}$, which may be much larger in a clean material at low temperature than the Planckian time $\tau_p$.  Then a local equilibrium can be reached via collisions between excitations without relaxation of momentum; a well-studied example motivating our study is with particle-hole symmetry~\cite{Damle1997,Sachdev1997,Sachdev1998}. If the conservation laws of the system such as energy, charge and momentum determine the relevant degrees of freedom beyond a certain time scale, one expects approximate phenomenological relativistic hydrodynamic equations to capture the coarse-grained properties~\cite{Landau2013,Kadanoff1963,Lucas_2018,Maxime2020} up to the momentum relaxation scale, and hydrodynamical effects will influence the measured conductivity.

A major experimental motivation for such models comes from the normal state of cuprate superconductors, and we summarize some of that very briefly.  While the origin of linear-in-temperature resistivity in the normal state of high $T_c$ superconductors at optimal doping remains an open question~\cite{Takagi1992,Zaanen2004,Zaane2019,Bruin2013,Varma_2020}, hydrodynamic studies for various kinds of quantum critical fluids suggest one kind of answer\cite{Ku2020,Muller20081,Muller20082,Bruin2013,Zaanen2004,Gallagher158,Nam2017,Stewart2001,Zaane2019}: a quantum critical electron fluid with maximal Planckian dissipation is one theoretical route to linear-$T$ resistivity, even if the nature of a quantum critical point near optimal doping is difficult to probe because of the intervening superconductivity.  The theoretical study of fast relaxation in higher dimensions was reinvigorated by the introduction of the Sachdev-Ye-Kitaev model, an analytically tractable nonlocal model of randomly interacting fermions~\cite{syk} with some unusual features such as ground-state entropy and all-to-all interactions that make its connection to materials somewhat opaque.

The Dirac fluid introduced above is a different route to linear-$T$ resistivity that is thought to be relevant to studies of transport in clean graphene samples near half-filling~\cite{Kimgraphene}.  
We can ask which quantum liquids in one dimension, where additional non-perturbative methods are available, can possess similar transport properties to charge-neutral graphene, and what the leading corrections to this behavior are in realistic systems.  It is well known that ballistic transport (i.e., free motion of carriers without scattering) in one dimension has quantized conductance at zero temperature.  Some special systems in one dimension, the integrable systems mentioned above, possess an extensive set of conservation laws that protect the current from relaxing~\cite{Lieb1963,Hild2014,Tang2018,Giamarchi2003quantum}, which often, though not always, leads to ballistic transport up to high temperature~\cite{Bertini_2021}.

Previous models have been studied to explore whether it is possible to relax the current in an impurity-free, non-integrable 1D system at finite temperature, but these generally have parametrically slower relaxation than required for linear-in-$T$ resistivity.  One example is to sit slightly away from  half-filling, with the band curvature considered and three-particle scattering processes introduced, which gives a quasi-particle decay rate scaling with the eighth power of the energy~\cite{Samokhin_1998,Matveev2013}.  Another example is to add staggered magnetic field at half-filling, to break the integrability of the 1D XXZ model, which turns out to give a power-law DC conductivity with respect to temperature depending on the Luttinger parameter~\cite{Huang2013,Bulchandani12713}.  These encourage us to look elsewhere for a 1D model that can support Planckian dissipation and linear-$T$ resistivity, in analogy with the Dirac fluid.

Our route is to use a scattering process in 1D, which can be thought of as either particle-hole scattering or fast Umklapp-like scattering (FUS), to approach limits on the relaxation of current in one dimension in an impurity-free system.  We start from a one-dimensional Dirac-like continuous model, where the Fermi level lies at the Dirac node (charge neutrality). The scattering will take two electrons from one linearly dispersing chiral branch to the opposite one, with no net momentum transferred. This is  different from the conventional Umklapp scattering in an ordinary 1D metal, where the two chiral branches are separated in momentum space at different Fermi points $k_L$ and $k_R$, such that the scattering process will possess a large momentum transfer $2|k_L-k_R| \sim 2\pi$. From the bosonization of the low energy Hamiltonian, we find that the FUS terms are irrelevant and that the Fermi liquid quasi-particle picture survives. The scattering can thus predominantly contribute to the current relaxation while adding only a small modification to the energy spectrum.  Our focus will be on realizing this mechanism while avoiding gap-opening instabilities such as dimerization and also Luttinger liquid behavior from forward scattering.

This scattering is thus an ``extrinsic'' mechanism~\cite{Else2021} in the sense that the fixed point is still a Fermi liquid, but the scattering can still be strong enough to give linear-in-$T$ resistivity.  With the quasi-particle picture preserved, combined with conservation of energy, momentum, and charge, we can then use a kinetic theory approach to calculate the DC conductivity. At the perturbative level, the collisionless limit (non-interacting case) gives a quantized conductance identical to that of ballistic transport with two chiral conducting channels.  In the presence of collisions, the conservation laws and particle-hole symmetry put a strong constraint on scattering processes. By using a standard variational method, we find that the conductivity is broadened to a Lorentzian by the FUS at finite temperature, and obtain a linear-$T$ resistivity in the DC limit.

In order to check that this physics can be realized in a solid, we then introduce a microscopic lattice model that manifests the aforementioned scattering process and transform it into a spin model via a Jordan-Wigner transformation. We use time-dependent density-matrix renormalization-group (DMRG) simulations~\cite{White_1992,Vidal2003,Vidal2007,Schollowock2011} to compute the current relaxation at finite temperature. Our results are consistent with the predictions of the field theory. Having a concrete model of linear-$T$ resistivity, in a numerically tractable system with local couplings and interactions, opens the door for future studies of the strong-coupling stability and generality of this kind of transport.

\section{Difference between the fast Umklapp scattering and conventional Umklapp scattering}
One crucial point for our paper is to tell the difference between two Umklapp processes: (1) fast Umklapp scattering process, as shown in Fig.~[Maintext.1c], (2) conventional Umklapp scattering process, as shown in Fig.~[Maintext.1d]. In either case, the decomposition for electron operator reads:
\begin{equation}
	\psi \approx e^{ +ik x} \psi_L + e^{-ik x} \psi_R,
\end{equation}
where $k$ is the momentum of corresponding excitation in the Brillouin zone. It can also be written in terms summation of Fermi point $k_F$ and the small deviation $p$ measured from the Fermi point, i.e., 
\begin{equation}
	k = k_F + p.
\end{equation}

The illustration for fast Umklapp scattering is shown in Fig.~[1c]. In this case, the fermi point sits at $k_F^0 = 0$. The momentum for two initial states reads:
\begin{equation}
	k_1 = p_1 + 0, \quad k_2 = p_2 +0.
\end{equation}
Here, $p_1$ and $p_2$ is the momentum measured from fermi point $0$. Similarly, the momentum for two final states reads:
\begin{equation}
	k_3 = p_3 + 0, \quad k_4 = p_4 + 0,
\end{equation}
where $p_3$ and $p_4$ is the momentum measured from fermi point $k_F^0 = 0$. Note that, in this case, the left Fermi point and right Fermi point coincides at the same point $k_F^0 = 0$. The total momentum change for the fast Umklapp scattering process reads:
\begin{equation}\label{FUS}
	\Delta k_f = k_1 + k_2 - k_3 - k_4 = p_1 + p_2 - p_3 - p_4.
\end{equation}

The illustration for conventional Umklapp scattering is shown in Fig.~[1d]. The momentum for two initial states reads:
\begin{equation}
	k_1 = p_1 + \pi/2, \quad k_2 = p_2 + \pi/2.
\end{equation}
Note that, $p_1$ and $p_2$ is the momentum measured from the right fermi point $k_F^+ = +\pi/2$, with $|p_1|,|p_2| \ll \pi/2$. Similarly, the momentum for two final states reads:
\begin{equation}
	k_3 = p_3 - \pi/2, \quad k_4 = p_4 - \pi/2,
\end{equation} 
where $p_3$ and $p_4$ is the momentum measured from the left fermi point $k_F^- = -\pi/2$, with $|p_3|, |p_4| \ll \pi/2$. The total momentum change for the conventional Umklapp scattering process reads:
\begin{equation}\label{CUS}
	\Delta k_c = k_1 + k_2 - k_3 - k_4 = p_1 + p_2 - p_3 - p_4 + 2\pi. 
\end{equation}

Comparing Eq.~[\ref{CUS}] with Eq.~[\ref{FUS}], we find the momentum transfer for conventional Umklapp scattering process $\Delta k_c$ is greater then that of the fast Umklapp scattering process $\Delta k_f$ by $2\pi$. Consider long range 2D Coulomb interaction which takes the form $V(q) \propto 1/|q|$, where $q$ is the momentum transfer in the scattering process, in general we have interaction strength for fast Umklapp $V(\Delta k_f)$ stronger than that of the conventional Umklapp process $V(\Delta k_c)$. The lattice model for this part is the non-interacting part ($\tilde H_0$) of Eq.~[Maintext.16].

\section{Bosonization}\label{BosonizationSec}
The standard bosonized Hamiltonian for Eq.~[MainText.4] reads~\cite{Giamarchi2003quantum,Xu2006,Wu2006} (technical details for bosonizations can be found in Sec.~[\ref{DetailsForBosonization}]):
\begin{equation}\label{BosonizeedHamiltonian}
	H_B = \int [dx] \bigg{\{}\frac{v}{2} \bigg{[} \frac{(\partial_x \phi)^2}{K} + K(\partial_x \theta)^2 \bigg{]} - \frac{V\cos [\sqrt{16\pi} \phi(x)]}{2(\pi \alpha)^2}  \bigg{\}},
\end{equation}
where $\phi(x) = \phi_R(x) + \phi_L(x)$ and $\theta(x) = \phi_R(x) - \phi_L(x)$ are linear combinations of the bosonic fields $\phi_{R,L}(x)$, and $\alpha$ stands for a short-range cut-off, say the scale of lattice constant. As there is no forward scattering in Eq.~[MainText.5], such that the interacting strength for forward scattering $V_{\rm fw} =0$. Thus for our model Eq.~[MainText.4], we have the renormalized Fermi velocity $v=\sqrt{v_F^2 - V_{\rm fw}^2} = v_F$, and the Luttinger parameter $K =\sqrt{(v_F - V_{\rm fw})/(v_F + V_{\rm fw})} = 1$. The renormalization group analysis~\cite{Giamarchi2003quantum,Wu2006} shows that the last term in Eq.~[\ref{BosonizeedHamiltonian}] (FUS) is irrelevant when $K > 1/2$, providing our system is still gapless in the weak interacting limit and can be captured by Fermi liquid theory with well-defined quasi particles.

\section{Collisionless transport}\label{Collisionless}
 We can use the standard equation of motion analysis to write down the collisionless transport equations for the excitations. We define the distribution functions under the energy basis $\gamma_\pm$ at time $t$:
\begin{equation}
	f_\lambda(k,t) = \langle \gamma^\dagger_\lambda(k,t) \gamma_\lambda(k,t) \rangle.
\end{equation}
In the equilibrium, without external perturbation, these are related to the Fermi distribution function $f^0(p) = 1/(e^{(p-\mu)/k_BT}+1)$, with $\mu$ the chemical potential, thus
\begin{equation}
	f_\pm(k,t) = f^0(\pm \epsilon_k)=\frac{1}{e^{ (\pm \epsilon_k -\mu)/k_BT}+ 1}.
\end{equation}
In the absence of interactions, we have linear dispersion $\epsilon_\lambda(k) =\lambda \epsilon_k =\lambda  v_F|k|$ from Eq.~[MainText.3], with $\lambda = \pm 1$ for the excitations with positive and negative energy, respectively. Here and later, for simplicity we set $\hbar = k_B = 1$, and will only reinsert them back when needed. To the zeroth order, in the presence of an external electric field $E(t)$, in the collisionless limit the dynamics is captured by the following simple kinetic equation:
\begin{equation}\label{KineticEquationSupp}
	\bigg{[} \frac{\partial}{\partial t} + Q E(t) \frac{\partial}{\partial k} \bigg{]}f_\lambda(k,t) = 0. 
\end{equation}
We seek to solve the above kinetic equation within the standard approximation of linear response. First, we parametrize the change in $f_\lambda$ from its equilibrium value by using the following ansatz~\cite{Fritz2008,Arnold_2000}:
\begin{equation}\label{AnsatzSupp}
\begin{aligned}
	f_\lambda({k},\omega) &= 2\pi \delta(\omega)f^0(\lambda \epsilon_k) +Q \frac{kE(\omega)}{|k|} f^{0}(\lambda \epsilon_k)\times [1-f^0(\lambda \epsilon_k)]g_\lambda(\epsilon_k,\omega),
\end{aligned}
\end{equation}
with $g_\lambda(\epsilon_k,\omega)$ a function to be determined. Note that, we have Fourier transformed time $t$ to the frequency domain $\omega$, and replaced $f_\lambda(k,t)$ with $f_\lambda(k,\omega)$. When $\mu =0$, the system is at particle-hole symmetric point, and an applied electric field $E(\omega)$ generates deviations in the distribution functions for particles and holes with opposite signs. This is due to the fact that the driving term Eq.~[\ref{KineticEquationSupp}] is odd under $\lambda \rightarrow -\lambda$, thus the deviation also has to be asymmetric in $\lambda$:
\begin{equation}\label{gfunction}
	g_\lambda(\epsilon_k,\omega) = \lambda g(k,\omega).
\end{equation}
In coordinate space, there will be newly generated holes (particles) moving align (anti-align) with the external electric field. This can be viewed as the generation of particle-hole pairs. For the states within the orange and purple square shown in the Fig.~[MainText.2.(a)], at the same $k$ point, the particle and hole has opposite momentum, and each particle-hole pair has zero total momentum defined by Eq.~[MainText.8] in the presence of particle-hole symmetry. On the other hand, since the particles and holes carry opposite charge, if they move in the opposite directions, the total current given by Eq.~[MainText.7] is non-zero.

Substituting Eq.~[\ref{AnsatzSupp}] into Eq.~[\ref{KineticEquationSupp}], one could derive a solution for $g$:
\begin{equation}\label{NonInteractingSolution}
	g_\lambda(\epsilon_k,\omega) = \frac{\lambda v_F /T}{-i\omega+ \eta},
\end{equation}
with $\eta$ is a positive infinitesimal. One can further insert this into the expression of current operator Eq.~[MainText.7.(b)] to get the conductivity: 
\begin{equation}\label{FreeConductivity}
	\begin{aligned}
		\sigma(\omega) &= \frac{\langle J \rangle}{E(\omega)} = Qv_F \sum_\lambda \int \frac{dk}{2\pi} \frac{\lambda k}{|k|}  \bigg{\{} Q \frac{k}{|k|} f^0(\lambda \epsilon_k)[1-f^0(\lambda \epsilon_k)]g_\lambda(\epsilon_k,\omega) \bigg{\}}\\
		&=\frac{2Q^2v_F^2/T}{(-i\omega + \eta)} \int_{-\infty}^{+\infty} \frac{dk}{2\pi} \frac{k^2}{|k|^2} \bigg{[} -T\frac{\partial f^0( \epsilon_k)}{\partial (\epsilon_k)} \bigg{]}  \approx \frac{2Q^2}{h} \frac{\hbar v_F }{-i\hbar \omega + \eta}.
	\end{aligned}
\end{equation}
In the first line, we have used the fact that contribution from the unperturbed distribution function (integral related to the first term in Eq.~[\ref{AnsatzSupp}]) should vanish. Note that we have restored $\hbar$ and $k_B$ in the last line from dimensional analysis, in the second line. We also have used the relation that $f^0(p)[1-f^0(p)]= -T\partial_p f^0(p)$ is an even function of $p$, and the extra factor of $2$ comes from the summation of particle and hole channels. This is consistent with the bosonization results for clean system in $1$D~\cite{Giamarchi2003quantum}. As the $v_F/\omega$ has the unit of length, this in accordance with the fact that the conductivity in 1D is roughly the conductance times the length. In the low frequency limit, the above result is reduced to the Drude peak, which is the signal for ballistic transport~\cite{Giamarchi2003quantum,Sirker2009}.

\section{Transport with collisions}

In this section, we will show the zero momentum mode can be relaxed by the internal collisions among excitations even without disorder and the possibility of localization. With the interaction given by Eq.~[MainText.5] or Eq.~[MainText.6], the corresponding collision terms can be derived by a simple application of Fermi's golden rule. From the bosonization results for continuous model showed in Sec.~[\ref{BosonizationSec}], one can safely assume that turning on the interaction will neither open a gap, nor have an immediate modification of single-particle spectrum of Eq.~[MainText.3], such that $\epsilon_\lambda(k) =\lambda v_F|k|$, with $\lambda = \pm$ for two flavors of quasiparticles: the positive energy ones with the distribution function $f_{+}(k,t)$, and the negative energy ones with the distribution function $f_-(k,t)$. However, when it turns to a lattice model, it will be crucial that we take steps to ensure that the interaction does not contain additional terms that might open a gap. We have implicitly assumed that there are no additional conservation laws, as would happen in the special case of integrable models.

From the above, we arrive at the quantum Boltzmann equation with collisions~\cite{Fritz2008,Arnold_2000,Sachdev1998}:
\begin{equation}\label{QBECollisionSupp}
	\begin{aligned}
		\bigg{[} \frac{\partial}{\partial t} + Q {E}(t)  \frac{\partial}{\partial { k}} \bigg{]}f_\lambda({ k}, t)  &= -\frac{2\pi}{v_F}  \int \frac{dk_1}{2\pi } \frac{dq }{2\pi} {\mathcal R}.
	\end{aligned}
\end{equation}
The integrand ${\mathcal R} = {\mathcal R}_1 + {\mathcal R}_2$, capturing the scattering among excitations, can be divided into two parts. The first part is the scattering among different flavors of excitations (particle-hole to particle-hole):
\begin{equation}
\begin{aligned}
	{\mathcal R}_1 &= \delta[{(|k|-|{k_1}|)}-{(|{{k} + { q}}| - |{{k}_1 -{q}})}]R_1({k},{k}_1,{q}) \\
	&\times \{ f_\lambda({ k},t) f_{-\lambda}({ k}_1,t)[1-f_\lambda({ k}+{ q},t)][1-f_{-\lambda}({k}_1-{q},t)]- [1-f_\lambda({ k},t)][1- f_{-\lambda}({ k}_1,t)]f_\lambda({ k}+{ q},t)f_{-\lambda}({ k}_1-{ q},t)]\}, 
\end{aligned}
\end{equation} 
with scattering amplitude
\begin{equation}
	{R}_{1}({ k},{ k}_1,{ q}) = 4|T_{+--+}({ k},{k}_1,{q}) - T_{+-+-}({k},{k}_1,-{k}-{q} + { k}_1)|^2.
\end{equation}
The second part captures the scattering among same flavor of excitations (particles to particles or holes to holes):
\begin{equation}
	\begin{aligned}
		{\mathcal R}_2  &= \delta[{(|k|+|{k_1}|)}-{(|{{ k} + { q}| + |{k}_1 -{ q}})|}]R_2({ k},{ k}_1,{q})\\
		&\times \{ f_\lambda({k},t) f_{\lambda}({ k}_1,t)[1-f_\lambda({ k}+{ q},t)][1-f_{\lambda}({ k}_1-{ q},t)]
		- [1-f_\lambda({ k},t)][1- f_{\lambda}({k}_1,t)]f_\lambda({ k}+{ q},t)f_{\lambda}({ k}_1-{ q},t)]\},
	\end{aligned}
\end{equation}
with scattering amplitude:
\begin{equation}
	R_2({ k},{ k}_1, { q}) = 2 |T_{++++}({ k}, {k}_1, { q}) - T_{++++}({ k}, {k}_1, { k}_1 - { k} - { q})|^2.
\end{equation}
Note that we only have two integrals and one delta function on the right hand side of Eq.~[\ref{QBECollisionSupp}], as the conservation of energy and conservation of momentum have the same requirement for linear dispersion in one dimension.

As the interaction is invariant under $\lambda \rightarrow -\lambda$, the Eq.~[\ref{gfunction}] still holds. With this, we substitute the ansatz Eq.~[\ref{AnsatzSupp}] into the Eq.~[MainText.10]. As in previous work \cite{Fritz2008,Arnold_2000}, the solution of $g_\lambda(\epsilon_k,\omega) = \lambda g(k,\omega)$ is the stationary point of the following function ${\mathcal Q}[g]$, (i.e. $\delta Q[g]/\delta[g] =0$). For simplicity, one can define $\tilde k = v_F k/T$.  Such that the ${\mathcal Q}[g]$ can be written as:
\begin{equation}\label{VariationOriginal}
	\begin{aligned}
		{\mathcal Q}[g] &= \frac{\pi T^2}{4v_F^3} \int \frac{d\tilde k}{2\pi} \frac{d \tilde k_1}{2\pi} \frac{d\tilde q}{2\pi}  \frac{\delta(|\tilde k|-|\tilde k_1| - |{\tilde k} + {\tilde q}| + |{\tilde k}_1 -{\tilde q}|)R_1({\tilde k},{ \tilde k}_1,{ \tilde q})}{(e^{-|\tilde k|}+1)(e^{|\tilde k_1|}+1)(e^{|{\tilde k}+{ \tilde q}|}+1)(e^{-|{ \tilde k}_1 -{ \tilde q}|}+1)} \\
		&\times {[} \vartheta (\tilde k) g(\tilde k,\omega)- \vartheta (\tilde k_1)g(\tilde k_1,\omega) -\vartheta (\tilde k + \tilde q)g(|\tilde k+\tilde q|,\omega)+\vartheta(\tilde {k}_1 - \tilde {q})g(|\tilde k_1-\tilde q|,\omega){]}^2  \\
		&+ \frac{\pi T^2}{4v_F^3} \int \frac{d\tilde k}{2\pi} \frac{d \tilde k_1}{2\pi} \frac{d\tilde q}{2\pi}  \frac{\delta(|\tilde k|+|\tilde k_1| - |{\tilde k} + {\tilde q}| - |{\tilde k}_1 -{\tilde q}|)R_2({\tilde k},{ \tilde k}_1,{ \tilde q})}{(e^{-|\tilde k|}+1)(e^{-|\tilde k_1|}+1)(e^{|{\tilde k}+{ \tilde q}|}+1)(e^{+|{ \tilde k}_1 -{ \tilde q}|}+1)}  \\
		&\times [\vartheta(\tilde k) g(\tilde k, \omega)+ \vartheta(\tilde {k}_1)g(\tilde k_1 , \omega)-\vartheta({\tilde k +\tilde q})g(|\tilde k+\tilde q|, \omega)-\vartheta{(\tilde {k}_1 - \tilde {q})}g(|\tilde k_1-\tilde q|, \omega)]^2 \\
		&+\frac{T}{v_F} \int \frac{d\tilde k}{2\pi} \frac{g(\tilde k, \omega)[-i\omega g(\tilde k,\omega)/2-v_F/ T]}{(e^{|\tilde k|}+1)(e^{-|\tilde k|} +1)}.
	\end{aligned}
\end{equation}

We seek to understand the integrals in ${\mathcal Q}[g]$. First, in the non-interacting limit $R_1(\tilde k, \tilde k_1, \tilde q) =R_2(\tilde k,\tilde k_1, \tilde q) =0$, we note that both the first and the second integral in Eq.~[\ref{VariationOriginal}] vanishes. The stationary relation $\delta Q[g]/\delta g =0$ gives the same solution $g_\lambda(\epsilon_k,\omega)$ as in Eq.~[\ref{NonInteractingSolution}], in accordance with our results for collisionless limit discussed in Sec.~[\ref{Collisionless}]. To relax the current, the collisions (interactions) should be introduced, and the summation of integrals associated with $R_1(\tilde k, \tilde k_1, \tilde q)$ and $R_2(\tilde k,\tilde k_1,\tilde q)$ in Eq.~[\ref{VariationOriginal}] should be non-zero. By dimension analysis, we have~\cite{Fritz2008} $g(k,\omega) \approx \frac{v_F}{T^2}C(\omega)$ with $C(\omega)$ a dimensionless function. With this, we can pull the $g(k,\omega)\approx v_FC[\omega]/T^2$ functions out from the square brackets, leaving the summation of $\vartheta$ functions inside. The conservation of energy (delta function in the integrand) and the structure of the perturbed distribution function (square of the summation of sign functions) put an important constraint on the integral. For the inter-flavor scattering, the integrand $\delta(|\tilde k|-|\tilde k_1| - |{\tilde k} + {\tilde q}| + |{\tilde k}_1 -{\tilde q}|)\times \big{[}\vartheta(\tilde k) -\vartheta(\tilde k_1) - \vartheta(\tilde k+\tilde q) + \vartheta(\tilde k_1-\tilde q) \big{]}^2$ only allows a specific type of non-vanishing solution: initial-particle hole pairs moving in the opposite direction ($\tilde k = - \tilde k_1$) and then bouncing back with respect to each other ($\tilde k >0$, $\tilde k_1 <0$, $\tilde k+\tilde q <0$, $\tilde k_1 -\tilde  q >0$,  or $\tilde k <0$, $\tilde k_1 >0$, $\tilde k+ \tilde q>0$, $\tilde k_1 -\tilde q<0$). Similarly, for the intra-flavor scattering, the integrand $\delta(|\tilde k|+|\tilde k_1| - |{\tilde k} + {\tilde q}| - |{\tilde k}_1 -{\tilde q}|)\times \big{[}\vartheta(\tilde k) +\vartheta(\tilde k_1) - \vartheta(\tilde k+\tilde q) - \vartheta(\tilde k_1-\tilde q) \big{]}^2$ does not have a non-vanishing solution; this is in accordance with the fact that the 1D chiral fluid can not be relaxed in the absence of impurities. With everything mentioned above, we can simplify Eq.~[\ref{VariationOriginal}] to:
\begin{equation}\label{VariationFinal}
	\begin{aligned}
	&	{\mathcal Q}[C(\omega)]=		\int \frac{d\tilde k}{2\pi} \frac{C(\omega)[-i\omega(v_F/T^2)C(\omega)/2-v_F/ T]/T}{(e^{|\tilde k|}+1)(e^{-|\tilde k|} +1)}+\int \frac{d\tilde k}{2\pi}  \frac{d\tilde q}{2\pi} \frac{2R_1({\tilde k},{ -\tilde k},{ \tilde q})C^2(\omega)/(v_F T^2)}{(e^{-|\tilde k|}+1)(e^{|\tilde k|}+1)(e^{|{\tilde k}+{ \tilde q}|}+1)(e^{-|{ \tilde k} +{ \tilde q}|}+1)}.  \\
	\end{aligned}
\end{equation}
For a given interaction $V(q)$, one can determine the value of $R_1(\tilde k, -\tilde k_1, \tilde q)$ based on Eq.~[MainText.6] and Eq.~[MainText.10]. After further performing the integral numerically, the function ${\mathcal Q}[g]$ has the following structure:
\begin{equation}
		{\mathcal Q}[C(\omega)] \simeq  \frac{v_F}{2T^2}\bigg{[} \kappa C^2(\omega) -i\frac{\omega}{T}{C^2(\omega)} -2C(\omega) \bigg{]},
\end{equation}
where the $\kappa$ is the numerical result from the the integral associated with $R_1(\tilde k, -\tilde k_1,\tilde q)$in Eq.~[\ref{VariationFinal}], depending on the explicit structure of $V(q)$:
\begin{equation}\label{NumericalCoefficient}
\kappa =\int \frac{d\tilde k}{2\pi} \frac{d \tilde q}{2\pi} \frac{4R_1(\tilde k, -\tilde k,\tilde q)/v_F^2}{(e^{-|\tilde k|} +1) (e^{|\tilde k|} +1)(e^{|\tilde k+\tilde q|} +1)(e^{-|\tilde k + \tilde q|} +1)}.
\end{equation}
 The integrand in Eq.~[\ref{NumericalCoefficient}] is non-negative everywhere. In fact, for usual $V(q)$ and $R_1(\tilde k, -\tilde k, \tilde q)$, the integrand is mostly positive on the 2D plane. As a result, 
the $\kappa$, as well as the correction to conductivity due to particle-hole scattering, should be non-zero. From this we can solve the $\delta {\mathcal Q}[g]/\delta g = \delta{\mathcal Q}[C(\omega)]/\delta C(\omega) =0$ as:
\begin{equation}\label{SolutionVariation}
	C(\omega) = \frac{1}{\kappa - i(\omega/T)}, ~~ g_\lambda(k,\omega) = \frac{\lambda v_F}{T^2}\frac{1}{\kappa - i(\omega/T)}.
\end{equation}
Take the Eq.~[\ref{SolutionVariation}] back into the Eq.~[\ref{AnsatzSupp}], we shall get the ansatz in the presence of collision. Substitute the ansatz with collision considered to the current $J$ defined in Eq.~[MainText.7b], by performing similar calculations as Eq.~[\ref{FreeConductivity}] in Sec.~[\ref{Collisionless}],  we obtain the conductivity in the presence of interactions:
\begin{equation}\label{CollisionConductivitySupp}
	\begin{aligned}
		\sigma(\omega) &= \frac{\langle J \rangle}{E(\omega)}  \approx \frac{2Q^2}{h} \frac{\hbar v_F}{-i\hbar \omega + \kappa k_BT}.
	\end{aligned}
\end{equation} 
Note that we have restored $\hbar$ and $k_B$ from dimensional analysis.
Compared with the low frequency diverging result for the collisionless limit in Eq.~[\ref{FreeConductivity}], the conductivity with collisions has some broadening at finite temperature. The conductivity Eq.~[\ref{CollisionConductivitySupp}] shows that the zero momentum mode can be relaxed solely by the momentum conserved internal scattering process among excitaions, i.e., the FUS. This physical picture is further plotted in Fig.~[MainText.2.(b-c)]. Take the inverse of Eq.~[\ref{CollisionConductivitySupp}], the resistivity in the DC limit shows the linear-$T$ dependence, i.e., the Planckian dissipation:
\begin{equation}
	\rho(\omega \rightarrow 0)\sim  \frac{\pi \kappa k_B}{Q^2 v_F} T = AT,
\end{equation}
with the coefficient $A = \pi \kappa k_B/ Q^2 v_F$. A one-dimensional Dirac system whose linear dispersion survives the interaction can be captured by a single model-dependent parameter, the Fermi velocity $v_F$. Combined with the temperature $T$, the only time scale in the continuous limit is the Planckian time $\tau_p = \hbar/k_BT$. Such a time scale gives the scattering rate for particle-hole excitations in an impurity-free Dirac system, and sets up an upper bound for the resistivity at finite temperature $\rho =A T$. The coefficient is $A \propto \kappa \propto R_1/v_F^2 \propto |V(q)/v_F|^2$, which shows that the resistivity is also positively related to the interaction strength in the perturbative region, in accordance with the previous results in wrapping graphene sheet to large-diameter metallic carbon nanotubes~\cite{Balents_1997}.

\section{Low energy Hamiltonian in the chiral basis}

Consider the lattice Hamiltonian (Eq.~[Maintext.16]):
\begin{equation}\label{NewLatticeModel}
	\begin{aligned}
	   \tilde H &= \tilde H_0 + \tilde  H_2 + \tilde H_3 \\
		\tilde H_0 &= +t \sum_i [-i\xi a_i^\dagger b_i + i \xi b_i^\dagger a_i -ib_i^\dagger a_{i+1} + ia^\dagger_{i+1}b_i] \\
		 \tilde H_2 &= \frac{V_2}{4}{ \sum_i\big{[}  (a_i^\dagger b_i - b^\dagger_i a_i)(a^\dagger_{i+1}b_{i+1}-b^\dagger_{i+1}a_{i+1})} + { (b^\dagger_i a_{i+1} - a^\dagger_{i+1}b_i)(b^\dagger_{i+1}a_{i+2} -a^\dagger_{i+2}b_{i+1})  }\big{]} \\
		\tilde H_3 &= \frac{V_3}{4} { \sum_i\big{[} (a^\dagger_i a_i -b^\dagger_i b_i)(a^\dagger_{i+1}a_{i+1}-b^\dagger_{i+1}b_{i+1}) } + {  (b_i^\dagger b_i -a^\dagger_{i+1}a_{i+1})(b^\dagger_{i+1}b_{i+1}-a^\dagger_{i+2}a_{i+2}) \big{]}}.
	\end{aligned}
\end{equation}
\begin{figure}[th]
\centering 
\includegraphics[width=0.8\columnwidth]{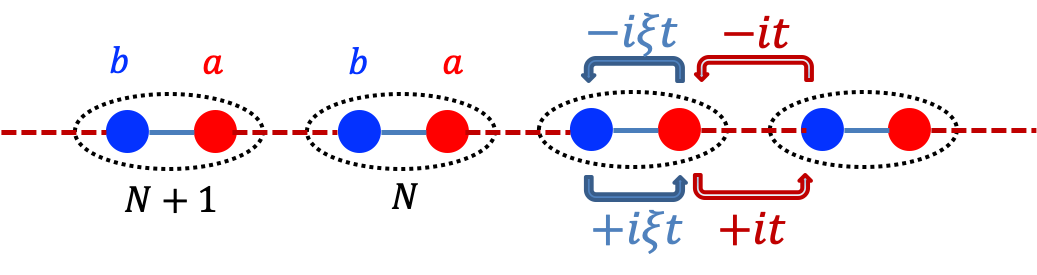}
\caption{\label{LatticeRealization}One realization of lattice model Eq.~[\ref{NewLatticeModel}]. The distance between two nearest neighbor sites is $a_0$, and the size for each unit cell is $2a_0$. Note that putting two sublattices at different points in the unit cell is merely for the convenience for pictorial illustration. Similar calculations can also be done for the case of two sublattices sitting in the same point in each unit cell, as described in the main text. The physics for these two cases are very similar.}
\end{figure}

\subsection{Non-interacting part}
We first write $\tilde H_0$ in the continuous limit at $\xi =1$:
\begin{equation}\label{FreeLattice}
	\begin{aligned}
		\tilde H_0 &= +t \sum_i [-i a_i^\dagger b_i + i b_i^\dagger a_i -ib_i^\dagger a_{i+1} + ia^\dagger_{i+1}b_i] \\
		&= + it \sum_i [ (a^\dagger_{i+1} -a_i^\dagger) b_i +  b_i^\dagger(a_i - a_{i+1}) ]
	\end{aligned}
\end{equation}
We define the slowly varying continuous fields $\psi_a(x)$ and $\psi_b(x)$ (this is possible, as our Bloch Hamiltonian as a Dirac cone at $k =0$, such that $k = k_F + p = p$ which is small and will only contribute a slowly varying phase factor $e^{ikx} \sim e^{ipx}$ compared with the scale of $a_0$ denotes distance between nearest neighbor of $a$ and $b$ sublattices), which satisfies~\cite{Fradkinbook}:
\begin{equation}\label{Continue}
	\psi_a[(2n-1)a_0] = \frac{1}{\sqrt{2a_0}}  a_n, \quad \psi_b[2na_0] = \frac{1}{\sqrt{2a_0}} b_n,
\end{equation}
such that we have:
\begin{equation}
	\begin{aligned}
		a^\dagger_{i+1} - a_i^\dagger \approx (2a_0)^{3/2} \partial_x\psi_a^\dagger (x), \quad a_i - a_{i+1} \approx -(2a_0)^{3/2} \partial_x \psi_a(x),
	\end{aligned}
\end{equation} 
where $x = 2ia_0$. Substitute this into Eq.~[\ref{FreeLattice}], we shall have the linearized non-interacting Hamiltonian:
\begin{equation}\label{Linearized}
	\begin{aligned}
		\tilde H_0 &= +i  (2a_0t) \int [dx] \{[\partial_x \psi_a^\dagger(x)]\psi_b(x)-\psi_b^\dagger(x) \partial_x \psi_a(x)\}, \\
		&= +i(2a_0t) \int [dx] [-\psi_a^\dagger(x) \partial_x \psi_b(x) - \psi_b^\dagger(x) \partial_x \psi_a(x)].
	\end{aligned}
\end{equation}
Now we adapt the following chiral basis to diagonalize the linearized Hamiltonian:
\begin{equation}\label{ChiralBasis}
	\psi_a(x) = \frac{1}{\sqrt{2}}[\psi_R(x) +\psi_L(x)], \quad \psi_b(x) = \frac{1}{\sqrt{2}} [\psi_R(x) - \psi_L(x)],
\end{equation} 
note that, this step is just changing basis and no approximation has been made. Substitute this back to Eq.~[\ref{Linearized}], we shall have:
\begin{equation}\label{NIntReal}
	\tilde H_0 = 2a_0 t \int [dx] [-i\psi^\dagger_R(x)\partial_x \psi_R(x) + i \psi_L^\dagger(x) \partial_x \psi_L(x)].
\end{equation}
Transform into momentum space, we will get the Eq. [Maintext.1]. 

\subsection{Interacting part}
Now combining Eq.~[\ref{Continue}] and Eq.~[\ref{ChiralBasis}]:
\begin{equation}
	\begin{aligned}
		a_n &= \sqrt{2a_0} \psi_a[(2n-1)a_0] = \frac{\sqrt{2a_0}}{\sqrt{2}} \{\psi_R[(2n-1)a_0] + \psi_L[(2n-1)a_0]\} \\
		b_n &= \sqrt{2a_0} \psi_b[2na_0] = \frac{\sqrt{2a_0}}{\sqrt{2}} \{ \psi_R[2na_0] - \psi_L[2na_0] \}.
	\end{aligned}
\end{equation}
As the continuous limit is taking $a_0 \rightarrow 0$, keeping the leading order of $a_0$, we shall have:
\begin{subequations}
	\begin{align}
		& a^\dagger_n b_n - b_n^\dagger a_n = 2a_0 [\psi_L^\dagger(x) \psi_R(x) - \psi_R^\dagger(x) \psi_L(x)] + {\mathcal O}(a_0^2) \\
		& a^\dagger_{n+1} b_n  - b^\dagger_n a_{n+1} =  2a_0 [\psi_L^\dagger(x) \psi_R(x) - \psi_R^\dagger(x) \psi_L(x)] + {\mathcal O}(a_0^2) \\
		& a_n^\dagger a_n - b^\dagger_n b_n = \psi_L^\dagger(x) \psi_R(x) + \psi_R^\dagger(x) \psi_L(x) + {\mathcal O}(a_0^2) \\
		& a^\dagger_{n+1} b_{n+1} - b^\dagger_{n+1} a_{n+1} = 2a_0 [\psi_L^\dagger(x+2a_0) \psi_R(x+2a_0) - \psi_R^\dagger(x+2a_0) \psi_L(x+2a_0)] + {\mathcal O}(a_0^2)\\
		& a^\dagger_{n+2} b_{n+1}  - b^\dagger_{n+1} a_{n+2}  = 2a_0 [\psi_L^\dagger(x+2a_0) \psi_R(x+2a_0) - \psi_R^\dagger(x+2a_0) \psi_L(x+2a_0)] + {\mathcal O}(a_0^2)
	\end{align}
\end{subequations}
%
%%\begin{subequations}
%%	\begin{align}
%%		&a^\dagger_n b_n - b_n^\dagger a_n \approx a^\dagger_{n+1} b_n  - b^\dagger_n a_{n+1} \approx 2a_0 [\psi_L^\dagger(x) \psi_R(x) - \psi_R^\dagger(x) \psi_L(x)], \\
%%		& a_n^\dagger a_n - b^\dagger_n b_n \approx \psi_L^\dagger(x) \psi_R(x) + \psi_R^\dagger(x) \psi_L(x), \\
%%		&a^\dagger_{n+1} b_{n+1} - b^\dagger_{n+1} a_{n+1} \approx a^\dagger_{n+2} b_{n+1}  - b^\dagger_{n+1} a_{n+2} \approx 2a_0 [\psi_L^\dagger(x+2a_0) \psi_R(x+2a_0) - \psi_R^\dagger(x+2a_0) \psi_L(x+2a_0)].
%%	\end{align}
%%\end{subequations}
With this, we shall have the $\tilde H_2$ under chiral basis (to the leading order of $a_0$) reads:
\begin{equation}\label{H2Chiral}
	\begin{aligned}
		&(a_i^\dagger b_i - b_i^\dagger a_i)(a_{i+1}^\dagger b_{i+1} - b_{i+1}^\dagger a_{i+1}) \\
		\approx &(2a_0)^2[\psi_L^\dagger(x) \psi_R(x) - \psi_R^\dagger(x) \psi_L(x)][\psi_L^\dagger(x+2a_0) \psi_R(x+2a_0) - \psi_R^\dagger(x+2a_0) \psi_L(x+2a_0)] \\
		=& (2a_0)^2 [\psi_L^\dagger(x) \psi_R(x) \psi_L^\dagger(x+2a_0) \psi_R(x+2a_0) + \psi_R^\dagger(x) \psi_L(x) \psi_R^\dagger(x+2a_0)\psi_L(x+2a_0)] \\
		-& (2a_0)^2 [\psi_R^\dagger(x) \psi_L(x) \psi_L^\dagger(x+2a_0) \psi_R(x+2a_0) + \psi_L^\dagger(x)\psi_R(x) \psi^\dagger_R(x+2a_)) \psi_L(x+2a_0)] \\
	& (b_i^\dagger a_{i+1} - a_{i+1}^\dagger b_i)(b^\dagger_{i+1} a_{i+2} - a^\dagger_{i+2} b_{i+1})  \\
		\approx &(2a_0)^2[\psi_L^\dagger(x) \psi_R(x) - \psi_R^\dagger(x) \psi_L(x)][\psi_L^\dagger(x+2a_0) \psi_R(x+2a_0) - \psi_R^\dagger(x+2a_0) \psi_L(x+2a_0)] \\
		=& (2a_0)^2 [\psi_L^\dagger(x) \psi_R(x) \psi_L^\dagger(x+2a_0) \psi_R(x+2a_0) + \psi_R^\dagger(x) \psi_L(x) \psi_R^\dagger(x+2a_0)\psi_L(x+2a_0)] \\
		-& (2a_0)^2 [\psi_R^\dagger(x) \psi_L(x) \psi_L^\dagger(x+2a_0) \psi_R(x+2a_0) + \psi_L^\dagger(x)\psi_R(x) \psi^\dagger_R(x+2a_)) \psi_L(x+2a_0)].
	\end{aligned}
\end{equation}
And similarly, we shall have the $\tilde H_3$ under chiral basis (also to the leading order of $a_0$):
\begin{equation}\label{H3Chiral}
	\begin{aligned}
		&(a_i^\dagger a_i - b_i^\dagger b_i)(a_{i+1}^\dagger a_{i+1} - b^\dagger_{i+1} b_{i+1})   \\
		\approx & (2a_0)^2[\psi_L^\dagger(x) \psi_R(x) + \psi_R^\dagger(x) \psi_L(x)][\psi_L^\dagger(x+2a_0) \psi_R(x+2a_0) + \psi_R^\dagger(x+2a_0) \psi_L(x+2a_0)] \\
		=& (2a_0)^2 [\psi_L^\dagger(x) \psi_R(x) \psi_L^\dagger(x+2a_0) \psi_R(x+2a_0) + \psi_R^\dagger(x) \psi_L(x) \psi_R^\dagger(x+2a_0) \psi_L(x+2a_0)] \\
		+& (2a_0)^2 [\psi_L^\dagger(x) \psi_R(x) \psi_R^\dagger(x+2a_0) \psi_L(x+2a_0) + \psi_R^\dagger(x) \psi_L(x) \psi_L^\dagger(x+2a_0)\psi_R(x+2a_0)] \\
		& (b^\dagger_i b_i - a_{i+1}^\dagger a_{i+1})(b^\dagger_{i+1} b_{i+1} - a^\dagger_{i+2} a_{i+2})  \\
		\approx & (2a_0)^2[\psi_L^\dagger(x) \psi_R(x) + \psi_R^\dagger(x) \psi_L(x)][\psi_L^\dagger(x+2a_0) \psi_R(x+2a_0) + \psi_R^\dagger(x+2a_0) \psi_L(x+2a_0)] \\
		=& (2a_0)^2 [\psi_L^\dagger(x) \psi_R(x) \psi_L^\dagger(x+2a_0) \psi_R(x+2a_0) + \psi_R^\dagger(x) \psi_L(x) \psi_R^\dagger(x+2a_0) \psi_L(x+2a_0)] \\
		+& (2a_0)^2 [\psi_L^\dagger(x) \psi_R(x) \psi_R^\dagger(x+2a_0) \psi_L(x+2a_0) + \psi_R^\dagger(x) \psi_L(x) \psi_L^\dagger(x+2a_0)\psi_R(x+2a_0)].
	\end{aligned}
\end{equation}
Collecting Eq.~[\ref{NewLatticeModel},\ref{H2Chiral},\ref{H3Chiral}], when $V_2 = V_3 = V$ we shall have:
\begin{equation}\label{HIntReal}
	\begin{aligned}
		\tilde H_{\rm int} = \tilde H_2 + \tilde H_3 &\approx  V a_0 \int [dx] [\psi_L^\dagger(x) \psi_R(x) \psi_L^\dagger(x+2a_0) \psi_R(x+2a_0) + \psi_R^\dagger(x) \psi_L(x) \psi^\dagger_R(x+2a_0) \psi_L(x+2a_0)].
	\end{aligned}
\end{equation}
Transform this into the Fourier space, we shall have:
\begin{equation}
	\begin{aligned}
		\tilde H_{\rm int} &= V a_0 \int [dx] [\psi_L^\dagger(x) \psi_R(x) \psi_L^\dagger(x+2a_0) \psi_R(x+2a_0) + \psi_R^\dagger(x) \psi_L(x) \psi^\dagger_R(x+2a_0) \psi_L(x+2a_0)] \\
		&=V a_0 \int [dx] \int \frac{dp_1}{2\pi} \frac{dp_2}{2\pi} \frac{dp_3}{2\pi} \frac{dp_4}{2\pi} [\psi_L^\dagger(p_1) \psi_R(p_2) \psi^\dagger_L(p_3) \psi_R(p_4)] e^{-ip_1x + ip_2 x - ip_3(x+2a_0) + ip_4(x+2a_0)} \\
		&+ Va_0 \int [dx] \int \frac{dk_2}{2\pi} \frac{dp_1}{2\pi} \frac{dp_4}{2\pi} \frac{dp_3}{2\pi}[\psi_R^\dagger(p_2) \psi_L(p_1) \psi_R^\dagger(p_4) \psi_L(p_3)] e^{-ip_2x + ip_1 x -ip_4(x+2a_0) + ip_3(x+2a_0)} \\
		&= Va_0 \int  \frac{dp_1}{2\pi} \frac{dp_2}{2\pi} \frac{dp_3}{2\pi} \frac{dp_4}{2\pi} [\psi_L^\dagger(p_1) \psi^\dagger_L(p_3) \psi_R(p_4)\psi_R(p_2) ] (2\pi)\delta(p_2-p_1 + p_4 - p_3) e^{i2a_0(p_4 - p_3)} \\
		&+ Va_0 \int \frac{dp_2}{2\pi} \frac{dp_1}{2\pi} \frac{dp_4}{2\pi} \frac{dp_3}{2\pi} [\psi_R^\dagger(p_2) \psi_R^\dagger(p_4) \psi_L(p_3)\psi_L(p_1) ] (2\pi) \delta(p_1 -p_2 + p_3 - p_4)e^{i2a_0(p_3 - p_4)}, 
	\end{aligned}
\end{equation}
where we have used the relation: $\int[dx] \exp[-ip_1x + ip_2x -ip_3 x+ ip_4 x] = (2\pi)\delta(p_2-p_1 + p_4 - p_3)$.
\begin{equation}
	\begin{aligned}
		\tilde H_{\rm int }&= Va_0 \int  \frac{dp_1}{2\pi} \frac{dp_2}{2\pi} \frac{dp_3}{2\pi} \frac{dp_4}{2\pi} [\psi_L^\dagger(p_1) \psi^\dagger_L(p_3) \psi_R(p_4)\psi_R(p_2) ] (2\pi)\delta(p_2-p_1 + p_4 - p_3) e^{i2a_0(p_4 - p_3)} \\
		&+ Va_0 \int \frac{dp_2}{2\pi} \frac{dp_1}{2\pi} \frac{dp_4}{2\pi} \frac{dp_3}{2\pi} [\psi_R^\dagger(p_2) \psi_R^\dagger(p_4) \psi_L(p_3)\psi_L(p_1) ] (2\pi) \delta(p_1 -p_2 + p_3 - p_4)e^{i2a_0(p_3 - p_4)} \\
		&= Va_0 \int \frac{dk_1}{2\pi} \frac{dk_2}{2\pi} \frac{dq}{2\pi} \psi_L^\dagger(k_1+q) \psi_L^\dagger(k_2-q) \psi_R(k_2) \psi_R(k_1)e^{+i2a_0 q} \\
		&+ Va_0 \int \frac{dk_1}{2\pi} \frac{dk_2}{2\pi} \frac{dq}{2\pi} \psi^\dagger_R(k_1 +q) \psi_R^\dagger(k_2-q) \psi_L(k_2) \psi_L(k_1) e^{+i2a_0 q},
	\end{aligned}
\end{equation}
where we have chosen $(p_1,p_2,p_3,p_4) = (k_1+q,k_1,k_2,k_2-q)$ in the simplification from first to the third line, and $(p_1,p_2,p_3,p_4)= (k_1,k_1+q,k_2,k_2-q)$ in the simplification from the second line to the forth line. Since $\tilde H_{\rm int}$ is hermitian, we have: $\tilde H_{\rm int} = (\tilde H_{\rm int} + \tilde H^\dagger_{\rm int })/2$, such that we have:
\begin{equation}
	\begin{aligned}
		\tilde H_{\rm int} &= Va_0 \int \frac{dk_1}{2\pi} \frac{dk_2}{2\pi} \frac{dq}{2\pi} \psi_L^\dagger(k_1+q) \psi_L^\dagger(k_2-q) \psi_R(k_2) \psi_R(k_1)e^{+i2a_0 q} \\
		&+ Va_0 \int \frac{dk_1}{2\pi} \frac{dk_2}{2\pi} \frac{dq}{2\pi} \psi^\dagger_R(k_1 +q) \psi_R^\dagger(k_2-q) \psi_L(k_2) \psi_L(k_1) e^{+i2a_0 q} \\
		&= \frac{Va_0}{2} \int \frac{dk_1}{2\pi} \frac{dk_2}{2\pi} \frac{dq}{2\pi}[\psi_L^\dagger(k_1+q) \psi_L^\dagger(k_2-q) \psi_R(k_2) \psi_R(k_1) e^{+i2a_0q}+ \psi_R^\dagger(k_1+q)\psi_R^\dagger(k_2-q)\psi_L(k_2)\psi_L(k_1) e^{+i2a_0q}] \\
		&+ \frac{Va_0}{2} \int \frac{dk_1}{2\pi} \frac{dk_2}{2\pi} \frac{dq}{2\pi}[\psi_R^\dagger(k_1) \psi_R^\dagger(k_2) \psi_L(k_2-q) \psi_L(k_1+q) e^{-i2a_0q}+ \psi_L^\dagger(k_1) \psi_L^\dagger(k_2) \psi_R(k_2-q) \psi_R(k_1+q)e^{-i2a_0q}] \\
			&= \frac{Va_0}{2} \int \frac{dk_1}{2\pi} \frac{dk_2}{2\pi} \frac{dq}{2\pi}[\psi_L^\dagger(k_1+q) \psi_L^\dagger(k_2-q) \psi_R(k_2) \psi_R(k_1) e^{+i2a_0q}+ \psi_R^\dagger(k_1+q)\psi_R^\dagger(k_2-q)\psi_L(k_2)\psi_L(k_1) e^{+i2a_0q}] \\
		&+ \frac{Va_0}{2} \int \frac{dk_1}{2\pi} \frac{dk_2}{2\pi} \frac{dq}{2\pi}[\psi_R^\dagger(k_2-q) \psi_R^\dagger(k_1+q) \psi_L(k_1) \psi_L(k_2) e^{-i2a_0q}+ \psi_L^\dagger(k_2-q) \psi_L^\dagger(k_1+q) \psi_R(k_1) \psi_R(k_2)e^{-i2a_0q}] \\
		&= \int \frac{dk_1}{2\pi} \frac{dk_2}{2\pi} \frac{dq}{2\pi} [Va_0 \cos(2qa_0) ][\psi_R^\dagger(k_1+q) \psi_R^\dagger(k_2-q) \psi_L(k_2) \psi_L(k_1) + \psi_L^\dagger(k_1+q)\psi_L^\dagger(k_2-q)\psi_R(k_2)\psi_R(k_1)].
	\end{aligned}
\end{equation}
Note that, from line 4 to line 6, we have changed the variables: $k_2-q \rightarrow k_1$ and $k_1 + q \rightarrow k_2$. In the last line, we used the relation: $(e^{+i2a_0qx}+e^{-i2a_0qx})/2 = \cos(2q a_0)$. By defining real function $V(q) = Va_0 \cos(2qa_0)$, we get Eq.~[Maintext.5]. Note that, we can show the Eq.~[Maintext.5] is hermitian:
\begin{equation}\label{InteractionChiralBasis1}
\begin{aligned}
	H_{\rm int} &= \int \frac{dk_1}{2\pi} \frac{dk_2}{2\pi} \frac{dq}{2\pi} V(q)[\psi^\dagger_R(k_1 + q) \psi^\dagger_R(k_2-q) \psi_L(k_2) \psi_L(k_1) + \psi^\dagger_L(k_1 + q) \psi^\dagger_L(k_2-q)\psi_R(k_2) \psi_R(k_1)],
\end{aligned}
\end{equation}
and
\begin{equation}
\begin{aligned}
	H_{\rm int}^\dagger &= \int \frac{dk_1}{2\pi} \frac{dk_2}{2\pi} \frac{dq}{2\pi} V(q)[\psi_L^\dagger(k_1) \psi_L^\dagger(k_2) \psi_R(k_2-q) \psi_R(k_1 + q) + \psi_R^\dagger(k_1) \psi_R(k_2) \psi_L(k_2-q) \psi_L(k_1+q)],
\end{aligned}
\end{equation}
by making the change of variables: $k_1 \rightarrow k_2 -q$ and $k_2 \rightarrow k_1 + q$, we shall have:
\begin{equation}\label{InteractionChiralBasis2}
\begin{aligned}
	H_{\rm int}^\dagger &= \int \frac{dk_1}{2\pi} \frac{dk_2}{2\pi} \frac{dq}{2\pi} V(q)[\psi_L^\dagger(k_2-q) \psi_L^\dagger(k_1+q) \psi_R(k_1) \psi_R(k_2) + \psi_R^\dagger(k_2-q) \psi_R(k_1+q) \psi_L(k_1) \psi_L(k_2)].
\end{aligned}
\end{equation}
Comparing Eq.~[\ref{InteractionChiralBasis1}] and Eq.~[\ref{InteractionChiralBasis2}], we shall have $H_{\rm int} = H_{\rm int}^\dagger$.

\subsection{More details on Bosonization}\label{DetailsForBosonization}
Combining Eq.~[\ref{NIntReal}] and Eq.~[\ref{HIntReal}], we have the Hamiltonian in the continuous limit reads:
\begin{equation}\label{CombinLowEnergy}
	\begin{aligned}
	H_{0,c} &= \int [dx] (-iv_F)[\psi^\dagger_R(x)\partial_x \psi_R(x)-\psi^\dagger_L(x)\partial_x\psi_L(x)], \quad v_F = td_0, \\
		H_{\rm int,c}&=V \int[dx][\psi^\dagger_{R}(x)\psi_{L}(x)\psi^\dagger_{R}(x+d_0)\psi_{L}(x+d_0)+\psi^\dagger_{L}(x)\psi_{R}(x)\psi^\dagger_{L}(x+d_0)\psi_{R}(x+d_0)],
	\end{aligned}
\end{equation}
where $d_0 = 2a_0$ is the lattice constant in Fig.~[\ref{LatticeRealization}]. Following the standard bosonization approach~\cite{Giamarchi2003quantum,ShankarBook}, we have  
\begin{equation}
	\psi_R(x) = \frac{1}{\sqrt{2\pi \alpha}} e^{+i \sqrt{4\pi} \phi_+(x)}, \quad \psi_L(x) = \frac{1}{\sqrt{2\pi \alpha}} e^{-i \sqrt{4\pi} \phi_-(x)},
\end{equation}
where $\alpha$ is a short distance cutoff. We further have the following commutation relations:
\begin{equation}\label{Commutation}
	\begin{aligned}
		[\phi_\pm(x),\phi_\pm(y)] = \pm \frac{i}{4}{\rm sgn}(x-y), ~~ [\phi_+(x),\phi_-(y)] = \frac{i}{4}, ~~ \phi(x) = \phi_+(x) +\phi_-(x), ~~ [\phi(x),\phi(x+d_0)] = 0.
	\end{aligned}
\end{equation}

The single particle part is bosonized via standard approach~\cite{Giamarchi2003quantum,ShankarBook,Fradkinbook}:
\begin{equation}
	H_{0}^B = \int[dx] \frac{v_F}{2} \bigg{[}(\partial_x \phi)^2 + (\partial_x \theta)^2\bigg{]},
\end{equation} 
with $\phi = \phi_+ + \phi_-$ and $\theta = \phi_- -\phi_+$.

Now we would love to bosonize the interactions:
\begin{equation}
	\begin{aligned}
		{\mathcal H}_{\rm int}^{B}=&V( \psi^\dagger_{R}(x)\psi_{L}(x)\psi^\dagger_{R}(x+d_0)\psi_{L}(x+d_0)+\psi^\dagger_{L}(x)\psi_{R}(x)\psi^\dagger_{L}(x+d_0)\psi_{R}(x+d_0) ) \\
		=& \frac{V}{(2\pi \alpha)^2} \bigg{[} e^{-i\sqrt{4\pi}\phi_+(x)}e^{-i\sqrt{4\pi}\phi_-(x)}e^{-i\sqrt{4\pi}\phi_+(x+d_0)}e^{-i\sqrt{4\pi}\phi_-(x+d_0)}  +{\rm h.c.}\bigg{]}.
	\end{aligned}
\end{equation}
By using the fact when $[A,B]$ is a constant, 
\begin{equation}
	e^A e^B = e^{A+B} e^{\frac{1}{2}[A,B]},
\end{equation}
and combined with the commutation relation Eq.~[\ref{Commutation}], we shall have:
\begin{equation}
	\begin{aligned}
		e^{-i\sqrt{4\pi}\phi_+(x)} e^{-i\sqrt{4\pi}\phi_-(x)} &= e^{-i\sqrt{4\pi} \phi(x)} e^{-\frac{4\pi}{2}[\phi_+(x),\phi_-(x)]} = e^{-i\sqrt{4\pi}\phi(x)} e^{-\frac{\pi}{2}i}.
%		e^{-i\sqrt{4\pi}\phi_+(x+a)}e^{-i\sqrt{4\pi}\phi_-(x+a)} &= e^{-i\sqrt{4\pi} \phi(x+a_0)} e^{-\frac{4\pi}{2}[\phi_+(x),\phi_-(x)]} =e^{-i\sqrt{4\pi} \phi(x+a)} e^{-\frac{\pi}{2}i}
	\end{aligned}
\end{equation}
Each four fermion term now is bosonized as:
\begin{equation}
	\begin{aligned}
	e^{-i\sqrt{4\pi}\phi_+(x)}e^{-i\sqrt{4\pi}\phi_-(x)}e^{-i\sqrt{4\pi}\phi_+(x+d_0)}e^{-i\sqrt{4\pi}\phi_-(x+d_0)}=	e^{-i\pi}e^{-i\sqrt{4\pi}\phi(x)} e^{-i\sqrt{4\pi}\phi(x+d_0)}.
	\end{aligned}
\end{equation}
Thus eventually we have:
\begin{equation}
	\begin{aligned}
		{\mathcal H}_{\rm int}^B=&V( \psi^\dagger_{R}(x)\psi_{L}(x)\psi^\dagger_{R}(x+d_0)\psi_{L}(x+d_0)+\psi^\dagger_{L}(x)\psi_{R}(x)\psi^\dagger_{L}(x+d_0)\psi_{R}(x+d_0) ) \\
		=& -\frac{V}{(2\pi \alpha)^2} 2\cos [\sqrt{4\pi}\phi(x)+\sqrt{4\pi}\phi(x+d_0)]. 
	\end{aligned}
\end{equation}
The final bosonized Hamiltonian for our model reads:
\begin{equation}\label{TotalBosonized}
	H^B = \int[dx] \frac{v_F}{2} \bigg{[}(\partial_x \phi)^2 + (\partial_x \theta)^2\bigg{]}  -\int [dx]\frac{V}{2(\pi \alpha)^2} \cos [\sqrt{4\pi}\phi(x)+\sqrt{4\pi}\phi(x+d_0)]. 
\end{equation}
This is just a Sine-Gordon model, very much close to the results from bosonize an XXZ model, and is identical to that of helical Luttinger liquid. Note that, the interaction part in Eq.~[\ref{CombinLowEnergy}] does not have forward scattering, i.e., $V_{\rm fw} = 0$. This corresponds to the first term of  Eq.~[\ref{BosonizeedHamiltonian}] in the case that $v = \sqrt{v_F^2 - V^2_{\rm fw}} =v_F$ and $K = \sqrt{(v_F - V_{\rm fw})/(V_F + V_{\rm fw})} =1$. Take $d_0 \rightarrow 0$, the second part of Eq.~[\ref{TotalBosonized}] is reduced to the second term of Eq.~[\ref{BosonizeedHamiltonian}].

%The Umklapp term at a single site indeed vanishes because of the Pauli principle, but there is a nonvanishing cosine we consider arising from neighboring sites.  A pedagogical discussion of this point is in the book by Giamarchi~\cite{Giamarchi2003quantum}. The interaction Eq.~[\ref{HIntReal}] is equivalent to Eq. [6.26] in Ref.~\cite{Giamarchi2003quantum} (note that our system, Fig.~[\ref{LatticeRealization}], has two sublattices sitting at different positions thus the lattice constant is 2$a_0$). The bosonized results for our model Eq. [S9] is very similar to that of the XXZ model treated by Giamarchi, whose Umklapp is irrelevant for $K > 1/2$.  We agree that the cosine term that survives is irrelevant in the RG sense, in accordance with the fact that fast Umklapp scattering will not open a gap, which has been confirmed by the DMRG results (see Sec.~[\ref{SecDMAG}]).

\section{Jordan Wigner transformation}

The model Eq.~[MainText.16] can also be transformed into a spin model in a finite length lattice with open boundary condition. Upon using the Jordan-Wigner transformation~\cite{Bahovadinov2019}: 
\begin{equation}
	S_{i,a}^+ = a_i^\dagger e^{i \pi \sum_{k <i} (a^\dagger_k a_k + b^\dagger_k b_k)}, \quad S_{i,b}^{+} =b_i^\dagger e^{i \pi \sum_{k <i}(a^\dagger_k a_k + b_k^\dagger b_k) + a_i^\dagger a_i}, \quad S^{z}_{i,a} = a^\dagger_i a_i - 1/2, \quad S^z_{i,b} = b_i^\dagger b_i - 1/2,
\end{equation}
we arrive at the following transformation dictionary: 
\begin{equation}
	a_i^\dagger b_i - b_i^\dagger a_i = S_{i,a}^{+} S_{i,b}^{-} - S_{i,a}^{-}S_{i,b}^{+}, \quad b_i^\dagger a_{i+1}-a^\dagger_{i+1}b_i = S_{i,b}^{+}S_{i+1,a}^{-} - S_{i,b}^{-}S_{i+1,a}^{+}.
\end{equation}
Substituting the above into Eq.~[16], we obtain the spin model:
\begin{equation}\label{Spin}
    H^S = H_0^S +  H_2^S +  H_3^S ,
\end{equation}
where each term is given by:
\begin{subequations}\label{SpinDetails}
\begin{align}
	&H_0^S = -\sum_{i=1}^N \bigg{[} it\xi \bigg{(} S_{i,a}^{+} S_{i,b}^{-} - S_{i,a}^{-}S_{i,b}^{+} \bigg{)}  + it \bigg{(} S_{i,b}^{+}S_{i+1,a}^{-} - S_{i,b}^{-}S_{i+1,a}^{+} \bigg{)}\bigg{]} \\
	& H_2^S= \frac{V_2}{4} \sum_i\bigg{[}(S_{i,a}^{+} S_{i,b}^{-} - S_{i,a}^{-}S_{i,b}^{+})(S_{i+1,a}^{+} S_{i+1,b}^{-} - S_{i+1,a}^{-}S_{i+1,b}^{+}) +( S_{i,b}^{+}S_{i+1,a}^{-} - S_{i,b}^{-}S_{i+1,a}^{+} 
)( S_{i+1,b}^{+}S_{i+2,a}^{-} - S_{i+1,b}^{-}S_{i+2,a}^{+} ) \bigg{]} \\
 &H_3^S = \frac{V_3}{4} \sum_i \bigg{[} (S_{i,a}^{z}-S_{i,b}^{z})(S_{i+1,a}^{z}-S_{i+1,b}^{z})+(S_{i+1,a}^{z}-S_{i,b}^{z})(S_{i+2,a}^{z}-S_{i+1,b}^{z})\bigg{]}.
\end{align}
\end{subequations}

\begin{figure}[th]
\centering 
\includegraphics[width=1\columnwidth]{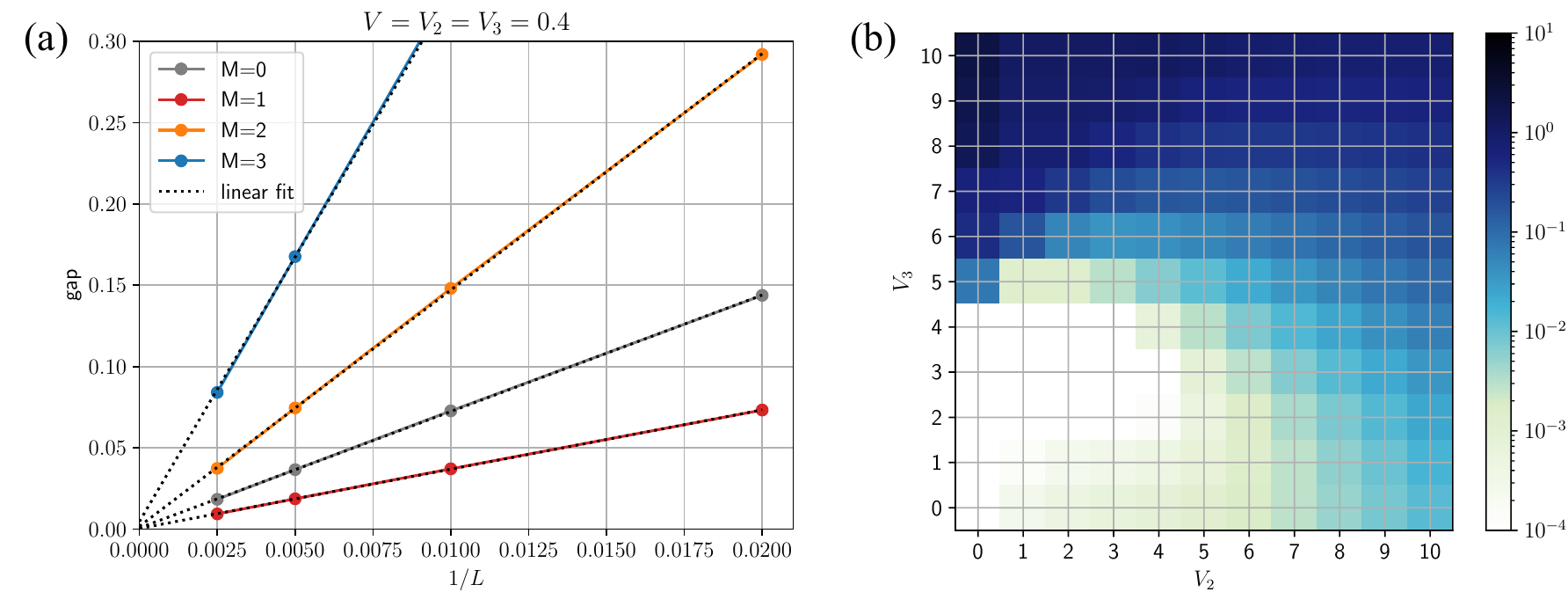}
\caption{\label{fig:gap_PD}(a) Excitation gaps in various sectors of the total magnetization $M$ of model Eq.~[16] or equivalently the Eq.~[\ref{SpinDetails}] for $V_2=V_3=0.4$ as a function of the inverse chain length $1/L$, calculated using DMRG with open boundary conditions. (b) Phase diagram of the excitation gap of model Eq.~[16] or equivalently the Eq.~[\ref{SpinDetails}] as a function of $V_2$ and $V_3$. For each point, the $M=1$ gap is interpolated in the system size for $L=100,200,400$.}
\end{figure}

\section{Connection to XXZ model and the breaking of integrablity}\label{ConnectionToXXZ}

Consider the following model, compared with Eq.~[\ref{SpinDetails}], we added one term $H_1^S$ (Eq.~[\ref{H1}]) controlled by parameter $\Delta$, 
\begin{subequations}\label{SpinDetailsNew}
\begin{align}
	&H_0^S = -\sum_{i=1}^N \bigg{[} it\xi \bigg{(} S_{i,a}^{+} S_{i,b}^{-} - S_{i,a}^{-}S_{i,b}^{+} \bigg{)}  + it \bigg{(} S_{i,b}^{+}S_{i+1,a}^{-} - S_{i,b}^{-}S_{i+1,a}^{+} \bigg{)}\bigg{]} \label{H0}\\
	&H_1^S = \Delta \sum_{i=1}^N  (S^z_{i,a} S^z_{i,b} + S^z_{i,b} S^z_{i+1,a}) \label{H1} \\
	& H_2^S= \frac{V_2}{4} \sum_i\bigg{[}(S_{i,a}^{+} S_{i,b}^{-} - S_{i,a}^{-}S_{i,b}^{+})(S_{i+1,a}^{+} S_{i+1,b}^{-} - S_{i+1,a}^{-}S_{i+1,b}^{+}) \nonumber \\
	&\quad \quad \quad \quad \quad +( S_{i,b}^{+}S_{i+1,a}^{-} - S_{i,b}^{-}S_{i+1,a}^{+} 
)( S_{i+1,b}^{+}S_{i+2,a}^{-} - S_{i+1,b}^{-}S_{i+2,a}^{+} ) \bigg{]} \label{H2} \\
 &H_3^S = \frac{V_3}{4} \sum_i \bigg{[} (S_{i,a}^{z}-S_{i,b}^{z})(S_{i+1,a}^{z}-S_{i+1,b}^{z})+{(S_{i+1,a}^{z}-S_{i,b}^{z})(S_{i+2,a}^{z}-S_{i+1,b}^{z})}\bigg{]} \label{H3}.
\end{align}
\end{subequations}
First, we will show that $H_0^S + H_1^S$ can be connected to XXZ model, in particular, $H_0^S$ corresponds to XX part and $H_1^S$ corresponds to ZZ part. 
For the $H_0^S$ part, we have:
\begin{equation}
	\begin{aligned}
		H_0^S =& - \sum_{i=1}^N \bigg{[} it\xi \bigg{(} S_{i,a}^+ S_{i,b}^-  - S^-_{i,a} S^+_{i,b} \bigg{)} + it \bigg{(} S^+_{i,b} S^-_{i+1,a} - S^-_{i,b} S^+_{i+1,a} \bigg{)} \bigg{]} \\
%		=& -it\xi  \sum_{i=1}^N [(S_{i,a}^x + i S_{i,a}^y)(S_{i,b}^x - iS_{i,b}^y) - (S_{i,a}^x - iS_{i,a}^y)(S_{i,b}^x + i S_{i,b}^y)] \\
%		&- it\sum_{i=1}^N [(S^x_{i,b} + i S^y_{i,b})(S^x_{i+1,a} - i S^y_{i+1,a}) - (S^x_{i,b} - iS^y_{i,b})(S^x_{i+1,a} + i S^y_{i+1,a})] \\
		=& 2 t \xi \sum_{i=1}^N [S^y_{i,a} S^x_{i,b} - S^x_{i,a}S^y_{i,b}] + 2t \sum_{i=1}^N [S^y_{i,b} S^x_{i+1,a} - S^x_{i,b} S^y_{i+1,a}].
	\end{aligned}
\end{equation}
Now we make the substitution:
\begin{equation}
	S_{i,a}^x \rightarrow (-1)^i\tilde S_{i,a}^x, \quad S_{i,a}^y \rightarrow (-1)^i \tilde S_{i,a}^x, \quad S_{i,a}^z \rightarrow \tilde S_{i,a}^z, \quad S_{i,b}^x \rightarrow \tilde S_{i,b}^y, \quad S_{i,b}^y \rightarrow - \tilde S_{i,b}^x, \quad S_{i,b}^z \rightarrow \tilde S_{i,b}^z,
\end{equation}
and they preserve the commutation relations:
\begin{equation}
\begin{aligned}
    &[\tilde S^x_{i,a}, \tilde S_{i,a}^y] = [(-1)^i S_{i,a}^x, (-1)^i S_{i,a}^y] = i S_{i,a}^z = i\tilde S_{i,a}^z,  \quad  [\tilde S^x_{i,b}, \tilde S^y_{i,b}] = [S_{i,b}^y,  - S_{i,b}^x] = i S_{i,b}^z = i \tilde S_{i,b}^z \\
	&[\tilde S^y_{i,a}, \tilde S^z_{i,a}] = [(-1)^i S_{i,a}^y, S^z_{i,a}] = i(-1)^i S^x_{i,a} = i \tilde S_{i,a}^x , \quad   [\tilde S^y_{i,b}, \tilde S_{i,b}^z] = [S^x_{i,b}, S_{i,b}^z] = - iS_{i,b}^y= i\tilde S_{i,b}^x, \\
	&[\tilde S^z_{i,a}, \tilde S^x_{i,a}] = [S_{i,a}^z, (-1)^uS^x_{i,a}] = i(-1)^i S^y_{i,a} = i \tilde S_{i,a}^y, \quad  [\tilde S^z_{i,b}, \tilde S_{i,b}^x] = [S_{i,b}^z,-S_{i,b}^y] = iS_{i,b}^x = i \tilde S^y_{i,b}.
\end{aligned}
\end{equation}
Under the new basis, we shall have:
\begin{equation}
\begin{aligned}
	H_0^S &= + 2t \xi \sum_{i=1}^N [S_{i,a}^y \tilde S_{i,b}^y + S_{i,a}^x \tilde S^x_{i,b}] + 2t \sum_{i=1}^N [-\tilde S_{i,b}^x S^x_{i+1,a} - \tilde S_{i,b}^y S_{i+1,a}^y ] \\
	&= - 2t \xi  \sum_{i=1}^N [\tilde S_{i,a}^y \tilde S_{i,b}^y + \tilde S_{i,a}^x \tilde S^x_{i,b}] - 2t \sum_{i=1}^N [\tilde S_{i,b}^x \tilde S^x_{i+1,a} + \tilde S_{i,b}^y \tilde S_{i+1,a}^y ]. 
\end{aligned}
\end{equation}
We choose $t = -1$ and $\xi = 1$, such that
\begin{equation}
	 H_0^S = 2 \sum_i^N [\tilde S_{i,a}^y \tilde S_{i,b}^y + \tilde S_{i,a}^x \tilde S^x_{i,b} +\tilde S_{i,b}^x \tilde S^x_{i+1,a} + \tilde S_{i,b}^y \tilde S_{i+1,a}^y  ] .
\end{equation}
If we set $\tilde S^{x,y,z}_{i,a} = \tilde S_{2i-1}$ and $\tilde S^{x,y,z}_{i,b} = \tilde S_{2i}$, we shall have $H_0^S$ reduced to a standard XX model:
\begin{equation}\label{XXComponent}
	 H_0^S = 2\sum_{n = 1}^{2N-1} [\tilde S_{n}^x \tilde S_{n+1}^x + \tilde S_n^y \tilde S_{n+1}^y].
\end{equation}
Similarly, we shall have the standard ZZ-component:
\begin{equation}\label{ZComponent}
	 H_1^S = \Delta \sum_{i=1}^N (S^z_{i,a} S^z_{i,b} + S^z_{i,b} S^z_{i+1,a}) = \Delta \sum_{i=1}^N (\tilde S^z_{i,a} \tilde S^z_{i,b} + \tilde S^z_{i,b} \tilde S^z_{i+1,a}) = \Delta \sum_{n=1}^{2N-1} \tilde S^z_{n} S^z_{n+1}.
\end{equation}
We further have $H_0^S +  H_1^S$ is connected to a standard integrable XXZ model:
\begin{equation}
	\tilde H_{\rm XXZ} = 2\sum_{n = 1}^{2N-1} [\tilde S_{n}^x \tilde S_{n+1}^x + \tilde S_n^y \tilde S_{n+1}^y + \Delta \tilde S^z_{n} \tilde S^z_{n+1} ].  
\end{equation}

\begin{figure}[!h]
\centering 
\includegraphics[width=1\columnwidth]{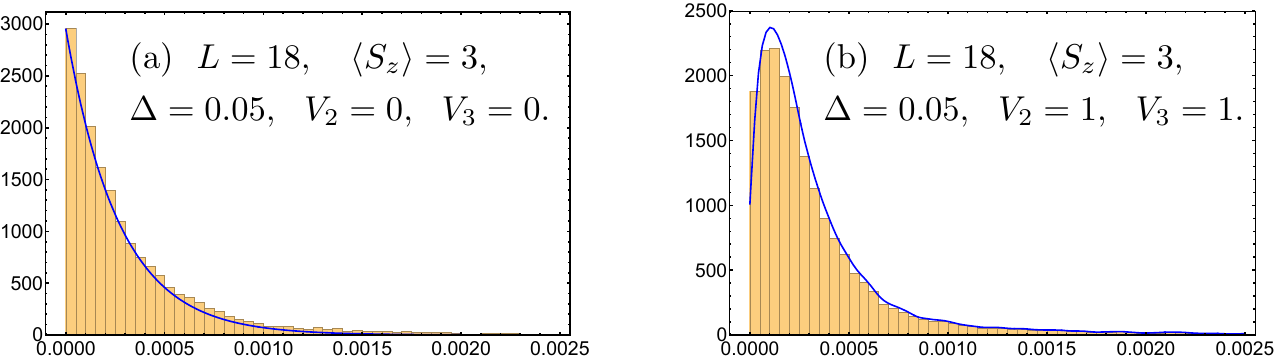}
\caption{\label{Picture1EXACT}Energy level spacing distribution for $t = -1$, $\xi = 1$, $\Delta = 0.05$, for a spin chain captured by Eq.~[\ref{SpinDetailsNew}] with 18 sites. We look into the $\langle S_z\rangle = 3$ section in open boundary condition. Note that we have introduced the finite $\Delta$ to avoid the collapse of energy level spacing distribution for XX model. (a)  Poisson distribution for $V_2 = V_3 =0$ in Eq.~[\ref{SpinDetailsNew}]. We have shown the connection between the $V_2= V_3 = 0$ case of Eq.~[\ref{SpinDetailsNew}] to a XXZ model in Sec.~[\ref{ConnectionToXXZ}]. (b) Wigner-Dyson distribution for $V_2 = V_3 = |t| =1 $ in Eq.~[\ref{SpinDetailsNew}].}
\end{figure}

\begin{figure}[!h]
\centering 
\includegraphics[width=1\columnwidth]{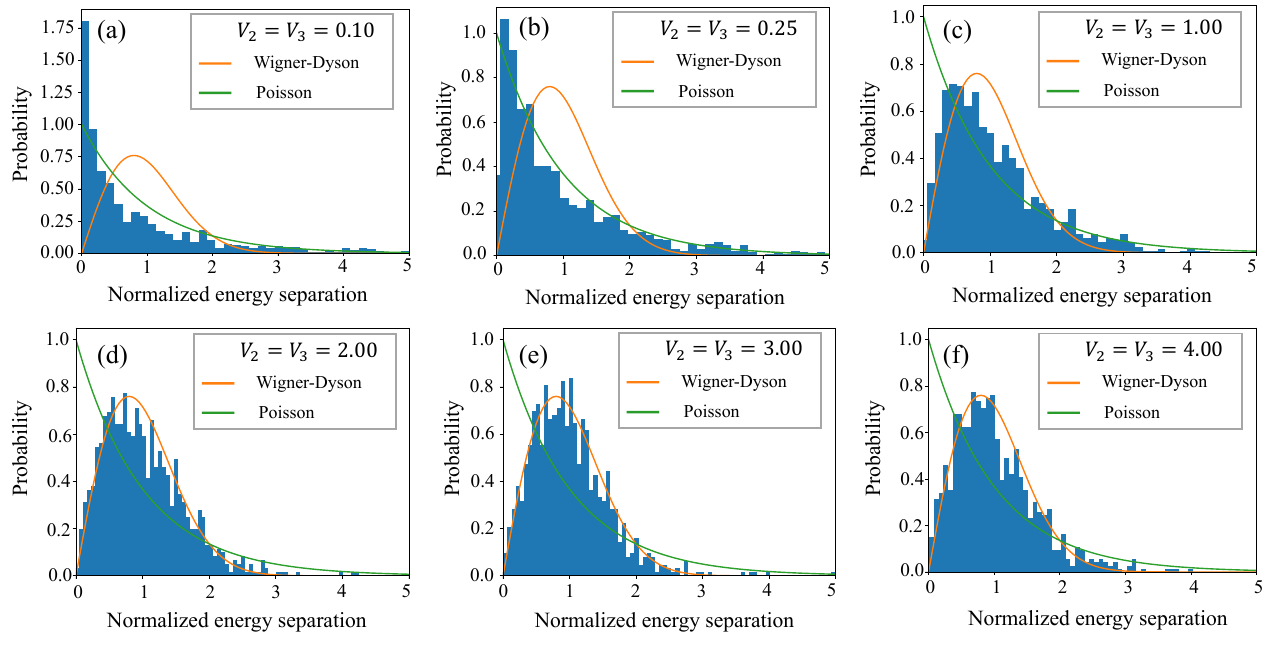}
\caption{\label{PBCLevelSpacing}Energy level spacing distribution for $t = -1$, $\xi =1$, $\Delta = 0$ with $V_2=V_3$ for different values in a spin chain given by Eq.~[\ref{SpinDetailsNew}] with $L = 18$ sites. We pick the spin sector $\langle S_z\rangle = 3$ and momentum sector $k = 2\pi/L$ in the presence of periodic boundary condition. We have dropped the case $V_2 = V_3 = 0$ where there are a lot of degeneracies. In this figure we follow the Gaussian broadening method to normalize the spectrum and make it comparable with probability distributions~\cite{Bruus1997}. We see that from $V_2 = V_3 = 0.10$ to $V_2 = V_3 = 4.00$, the energy level spacing distribution changes from Poisson distribution to Wigner-Dyson distribution. Both Fig.~[\ref{Picture1EXACT}] and Fig.~[\ref{PBCLevelSpacing}] gives the similar results as the Fig.~[2] in Ref.~\cite{Huang2013}.} 
\end{figure}

With above, we have shown that $H_0^S + H_1^S$ can be connected to XXZ model. The XXZ model is an integrable model, whose energy level spacing has a Poisson distribution. However, in the Fermionic picture, $H_0^S$ itself corresponds to a non-interacting fermionic model with approximate Dirac dispersion, and has lot's of degeneracy, making a lot of extra zeros in the energy level spacing distribution. To avoid this behavior and see a clear distribution function, we added the additional $H_1^S$ with small coefficient $\Delta$ which corresponds to the interaction term in the fermionic picture. The additional of small coefficient $H_1^S$ will neither break the integrability nor gap out $H_0^S$. We show that for open boundary condition, the energy level spacing of $H_0^S + H_1^S$ has a Poisson distribution, as shown in Fig.~[\ref{Picture1EXACT}.(a)]. On the other hand, 
both $H_2^S$ and $H_3^S$ breaks the integrability, as the energy level spacing of $H_0^S + H_1^S + H_2^S + H_3^S$ has the form of Wigner-Dyson distribution, see in Fig.~[\ref{Picture1EXACT}.(b)]. Similar result for periodic boundary condition has also been seen in Fig.~[\ref{PBCLevelSpacing}].

\section{DMRG Details}\label{SecDMAG}

At first, we confirm that the lattice model Eq.~[MainText.16] or equivalently, Eq.~[\ref{SpinDetails}] is gapless using the density matrix renormalization group~\cite{White_1992}, which expresses the the wavefunction as a variational matrix-product state~\cite{Schollowock2011}. The Hamiltonian contains fairly complicated three-site interaction terms, making it more complicated than typical tight-binding chains. 
By employing a general technique to represent the Hamiltonian as a matrix-product operator (MPO)~\cite{Hubig_2017}, we find that it can be achieved with an MPO size of $10\times 10$.
In the spin language, the model possesses a U(1) symmetry corresponding to the conservation of the total magnetization $M=\sum_i\langle S^z_i \rangle$ (equivalent to the particle number in the fermionic language), which is exploited in the DMRG algorithm.

The ground state is found in the $M=0$ sector (or half filling). We can look at the neutral gap $\Delta_0 = E_1(M=0)-E_0(M=0)$, as well as at the gaps in various magnetization sectors: $\Delta_{M_0} = E_0(M=M_0)-E_0(M=0)$ ($M_0>0$), which corresponds to the flipping of $M_0$ spins (or to the removal of $M_0$ particles in the fermionic language). Using DMRG, we compute these gaps for system sizes up to $L=400$. Fig.~[\ref{fig:gap_PD}.(a)] shows a typical result for $V=V_2=V_3=0.4$. We find that all gaps scale linearly with $1/L$, making an extrapolation for $L\to\infty$ very easy. The extrapolated values are very small (of the order of $10^{-4}-10^{-3}$), consistent with the expectation that the system remains gapless for finite $V$. We also calculate a full phase diagram in the $V_2-V_3$ plane, shown in Fig.~[\ref{fig:gap_PD}.(b)].  This gives us an indication of how $V_2$ and $V_3$ may be chosen while keeping the system gapless.

%{\color{blue} In the calculation of the gap, the maximal bond dimension for the gap is about $\chi = 300$. One can also assess the quality of the energies obtained by computing the variance per site, $E_v$ as,
%\begin{equation}
%	E_v = \frac{\langle  H^2\rangle - \langle H\rangle^2}{L}.
%\end{equation}
%The smaller the $E_v$ is, the closer one to an eigenstate and the better energy. We checked that in our case, $E_v < 10^{-7}$, which shows that our system is indeed gapless in the relevant region. }

{\color{black} The maximal bond dimension used for the gap is about $\chi = 300$, which is good enough for making a phase diagram.  A check is the variance per site of the ground state energy, $(<H^2> - <H>^2)/L$. The smaller it is, the closer one is to an eigenstate and the better the energy, and we checked that this quantity is always less than $10^{-6}$ in units where the energy scale is 1. The bond dimension used for the dynamical calculations is larger, $\chi =800$.}

During the time propagation for calculating the dynamics, the entanglement entropy grows, and we dynamically increase the bond dimension $\chi$ of the MPS representation by keeping the truncation error fixed; we have varied this control parameter in order to ensure that our results are converged. The imaginary time evolution is stopped once $\chi=800$, which allows us to reach temperatures of $1/T=16$. In order to reach lower $T$, we continue the cooling using a constant $\chi$, and we have verified that this second truncation parameter does not influence the results. While the truncation error is not fixed during the cooling beyond $1/T=16$, this regime is not key to our message, and the results should only be viewed as providing additional support. The ensuing real time evolution is stopped once $\chi=1600$ is reached, since any further propagation becomes prohibitively slow. We employ further standard optimizations that allow us to maximize the use of numerical resources~\cite{Kennes_2016}: (i) Counterpropagating the auxiliary space limits the growth of entanglement; (ii) Splitting up the propagation into a forward and a backward one increases the achievable $t_{\text{max}}$. In the forward propagation, we can (iii) exploit the approximate translational invariance in the middle of the chain (reducing the perturbation to the local current $j_{i+1}$), as well as (iv) the spinflip symmetry of the current operator.

\section{Current operator}
 The current density operator $j_{i+1}=tQ[b_i^\dagger a_{i+1} + a^\dagger_{i+1} b_i]$ defined in the main text  will be slightly modified if we wish to have a local continuity equation for every site, rather than every unit cell, as might be appropriate if the $a$ and $b$ sites are in different spatial locations, but this does not significantly affect the long-wavelength currents relevant for transport, as we have verified numerically in test cases. {In fact, we can simply define a system with no-internal spatial structure of each unit cell, and view the $a^\dagger_i$ ($b^\dagger_i$) as pseudo spin $c^\dagger_{\uparrow,i}$ ($c^\dagger_{\downarrow,i}$) sits right on the center of $i$-th unit cell. In the latter case, we only have one site per unit cell, and the operator  $\rho_n = Q(a^\dagger_n a_n + b^\dagger_n b_n) = Q(c^\dagger_{\uparrow,n}c_{\uparrow,n} + c^\dagger_{\downarrow,n} c_{\downarrow,n})$ precisely gives the $U(1)$ charge density on each site, as well as the corresponding unit cell.} Second,  with the symmetrized interaction we have chosen on the lattice, which is not a pure density-density interaction but rather only part of one, the total charge is clearly still conserved.  However, the equation of continuity is modified {by some terms beyond the leading order that we are interested in}. {In the lattice, the full current density that satisfies the continuity equation with local charge density $\rho_j = a^\dagger_j a_j + b^\dagger_j b_j$ would be: $j_j = j_j^0 + j_j^\prime$, with $j_j^0=tQ(b^\dagger_{j-1}a_j + a^\dagger_j b_{j-1}) $ is what we defined above, the higher order additional four-fermion terms $j_j^\prime = {(iV_2Q)}(b^\dagger_{j-2} a_{j-1}b^\dagger_{j-1} a_j- a^\dagger_{j-1} b_{j-2} a^\dagger_j b_{j-1} - a_{j-1}^\dagger b_{j-2}b^\dagger_{j-1} a_j + b^\dagger_{j-2}a_{j-1}a^\dagger_j b_{j-1} )/4+{(iV_2Q)}(b^\dagger_{j-1} a_{j}b^\dagger_{j} a_{j+1}- a^\dagger_{j} b_{j-1} a^\dagger_{j+1} b_{j} + a_{j}^\dagger b_{j-1}b^\dagger_{j} a_{j+1} - b^\dagger_{j-1}a_{j}a^\dagger_{j+1} b_{j} )/4$ comes from the interaction $V_2$ which is not in the form of product of density.}

\section{Discussion}

While the possibility of metals with linear-in-$T$ resistivity has been an actively discussed topic for many years, there are relatively few concrete models verified to possess this property, especially if one requires the interactions to be local and non-random.  The main results here start from developing a low-energy model for a 1D Dirac fermionic system in which particle-hole Umklapp-like scattering is the dominant process.  By using the standard bosonization theory, we showed that this interaction is irrelevant and will not modify the band structure to the leading order. We used kinetic equations to study the transport properties for a fermionic model with Dirac-like dispersion, which coincides with the well-known results for 1D two-channel ballistic transport. We further calculated the conductivity in the presence of the scattering, and show the broadening from collision of quasi-particles in finite temperature.

In the low frequency limit, the resistivity linearly depends on temperature, the feature known as Planckian dissipation.  
We further provided a lattice realization with the Dirac model as its low-energy limit.  By using the Jordan-Wigner transformation, we transformed the lattice model to a spin model. We were able to solve it for system sizes up to $L=400$ in the static case and up to $L=96$ in the dynamic case with finite temperature, by using the density-matrix renormalization group (DMRG). The results of the simulations are consistent with the predictions of the field theory. The scaling regime in this Dirac-like 1D fermionic model could be relevant in single-walled carbon nanotubes or cold atomic systems~\cite{Balents_1997,Yoshioka1999}, providing a route if the interactions are strong to the experimental observation of Planckian dissipation in 1D systems.

Verifying transport similar to that proposed for the 2D Dirac liquid in a 1D model provides an alternative point of view on the origin of bad metallic behavior, distinct from scenarios involving quantum criticality.  While one could view the Fermi gas as a point in the phase diagram, rather than a phase as in higher dimensions, observables generally evolve continuously into the Luttinger liquid, unlike moving from a quantum critical point into a neighboring phase.  Observing the dominance of Umklapp-like scattering in this model complements other possibilities for transport theory in one dimension dominated by other irrelevant operators~\cite{Rice2017}.

One direction for future work comes from isolating the relaxation time from other pieces of the conductivity to see whether there is a crossover with temperature in the source of the linear-in-temperature behavior.  This would allow comparison to such a crossover in conductivity of doped Hubbard models observed in recent work using quantum Monte Carlo continued to real time.\cite{Edwin2019} This would also allow direct comparison of the current relaxation time at strong interactions to the Planckian scale $\hbar / k_B T$.  Of course underlying physics in that study is almost certainly different, and our model is less directly relevant to the linear resistivity of high-$T_c$ superconductors.

Although we have focused in this work on the case where the system remaining gapless after turning on the interaction, one can generalize our treatment for the collision-dominated regime to gapped phases so long as the Landau Fermi-liquid quasi-particle description is still valid. The seemingly complicated model Eq.~[Maintext.16] provides a route to realizing the conjectured Planckian upper bound to the resistivity for a class of interacting semimetals in 1D.  More generally, the analytical and numerical methods available to explore transport in low spatial dimensions make it feasible to search for evidence of other physics originally proposed for higher dimensions, as we have done here for the Dirac fluid.

\end{document}